\documentclass[11pt,oneside,letter]{article}
\usepackage[T1]{fontenc}

\usepackage{amsthm, amsmath, amssymb, amsfonts, graphicx, epsfig, bbm}
\usepackage{mathrsfs}
\usepackage{enumerate}
\usepackage[colorlinks=true, pdfstartview=FitV, linkcolor=blue,
            citecolor=blue, urlcolor=blue]{hyperref}
\usepackage[usenames]{color}
\usepackage[numbers]{natbib}
\usepackage{url}
\makeatletter\def\url@leostyle{%
 \@ifundefined{selectfont}{\def\UrlFont{\sf}}{\def\UrlFont{\scriptsize\ttfamily}}} \makeatother
\usepackage{accents, comment}

\usepackage[normalem]{ulem}

\newtheorem{theorem}{Theorem}
\newtheorem{proposition}[theorem]{Proposition}

\newtheorem{corollary}[theorem]{Corollary}

\theoremstyle{definition}
\newtheorem{definition}[theorem]{Definition}
\newtheorem{example}[theorem]{Example}
\theoremstyle{remark}
\newtheorem{remark}[theorem]{Remark}

\numberwithin{equation}{section}
\numberwithin{theorem}{section}
\def\cB{\mathcal{B}}

\def\cF{\mathcal{F}}
\def\cG{\mathcal{G}}
\def\cH{\mathcal{H}}

\def\cK{\mathcal{K}}

\def\cX{\mathcal{X}}

\def\bF{\mathbb{F}}

\def\bN{\mathbb{N}}

\def\bR{\mathbb{R}}

\def\bT{\mathbb{T}}

\def\bV{\mathbb{V}}

\newcommand{\1}{\mathrm{I}}

\newcommand{\set}[1]{\{#1\}}            % set: {xyz} to be used for inline formulas
\newcommand{\Set}[1]{\left\{#1\right\}} % set: {xyz} to be used for seapare (not inline) formulas
\renewcommand{\mid}{\;|\;}              % mid bar with small spaces before and after: x | y
\DeclareMathOperator*{\esssup}{ess\,sup} % ess sup
\DeclareMathOperator*{\essinf}{ess\,inf} % ess inf

\title{Dynamic Limit Growth Indices in Discrete Time}

\author{Tomasz R. Bielecki$^{1}$\\[-0.3ex]
\url{bielecki@iit.edu} \\[-0.3ex]
\and
Igor Cialenco$^{1}$\\[-0.3ex]
\url{igor@math.iit.edu} \\[-0.3ex]
\and
Marcin Pitera$^{2}$\\[-0.3ex]
\url{marcin.pitera@im.uj.edu.pl}\\[-0.3ex]
\and
\small{$^{1}$Department of Applied Mathematics,}\\[-0.3ex]
\small{Illinois Institute of Technology,}\\[-0.3ex]
\small{Chicago, 60616 IL, USA}\\[-0.3ex]
\and
\small{$^{2}$Institute of Mathematics,}\\[-0.3ex]
\small{Jagiellonian University,}\\[-0.3ex]
\small{Cracow, Poland}\\[-0.3ex]
}
\date{First Circulated: December 03, 2013\\
This version: July 15, 2014}

\binoppenalty=\maxdimen %no breaks inside math formulas..
\relpenalty=\maxdimen

\begin{document}
\maketitle
\begin{abstract}
\noindent We propose a new class of mappings, called Dynamic Limit Growth Indices, that are designed to measure the long-run performance of a financial portfolio in discrete time setup.
We study various important properties for this new class of measures, and in particular, we provide necessary and sufficient condition for a Dynamic Limit Growth Index to be a dynamic assessment index.
 We also establish their connection with classical dynamic acceptability indices, and we show how to construct examples of Dynamic Limit Growth Indices using dynamic risk measures and dynamic certainty equivalents.
 Finally, we propose a new definition of time consistency, suitable for these indices, and we study time consistency for the most notable representative of this class -- the dynamic analog of risk sensitive criterion. \\
\\
{\noindent \small
 }
\end{abstract}

%\tableofcontents
%\newpage
\section{Introduction} %\addcontentsline{toc}{section}{Introduction}

In this paper we study some mappings that are designed to measure the long-run performance of a financial portfolio. Importance of measurement of the long run  growth of a portfolio is widely recognized among financial practitioners, and has been extensively discussed in the literature ( see for instance \cite{AraBorFerGhoMar1993, FleShe2000}, and references therein).

Here, we shall focus on measures that quantify the tradeoff between portfolio growth and the risk associated with it, appropriately normalized in time. Among several such possible measures, the one which has attracted the most attention, is the so called Risk Sensitive Criterion~\cite{Whi1990,BiePli1999,BiePli2003}.

In fact, the starting point of this paper was to investigate whether the Risk Sensitive Criterion belongs to the family of so called dynamic acceptability indices~\cite{CheMad2009,BieCiaZha2012}, which are known to provide a unifying framework for classical financial measures of performance such as Sharpe Ratio, Gain to Loss Ratio, etc.
It turns out that indeed the Risk Sensitive Criterion is a dynamic acceptability index. But, this investigation also led us to introducing a new class of mappings designed to measure the efficiency of the long run cumulative growth of a portfolio, which we named Dynamic Limit Growth Indices. Since, in this paper, we measure time according to discrete quanta, we only consider here dynamic limit growth indices in discrete time.

In some of the previous works studying the dynamic acceptability indices~\cite{CheMad2009,BieCiaZha2012}, the so called normalization postulate was adopted. Here, we remove the normalization postulate, thereby opening a doorway to a much richer class of dynamic performance measures, such as the class of our  dynamic limit growth indices.

In this paper we study all sorts of important properties for this new class. In particular, we provide necessary and sufficient condition for Dynamic Limit Growth Index to be a dynamic assessment index (cf. \cite{BieCiaDraKar2013}), we study their connection with classical dynamic acceptability indices, we show how to construct examples of dynamic limit growth indices using dynamic risk measures and dynamic certainty equivalents~\cite{CheKup2009}, we propose and study a definition of time consistency.

The paper is organized as follows. In Section~\ref{S:Preliminaries} we provide a set of some underlying concepts that will be used throughout the paper. In Section~\ref{S:DLGI} we introduce the notion of Dynamic Limit Growth Indices (DLGI), which is the main object of study of this work. Also here, we present necessary and sufficient conditions for a DLGI to be a dynamic assessment indices. Next we show a connection between dynamic limit growth indices and dynamic acceptability indices. We conclude the section by providing several classes of functions that are DLGI. In Section~\ref{S:time} we study the time consistency of DLGI. We propose a new definition of time consistency, and relate it to the existing literature. Finally, in Section~\ref{S:DRSC} we study into details the dynamic risk sensitive criterion. We prove that the dynamic risk sensitive criterion a  dynamic assessment index, and study its time consistency with respect to risk-sensitivity parameter. All the proofs are deferred to the Appendix.

\section{Preliminaries}\label{S:Preliminaries}
Let $(\Omega,\mathcal{F},\bF=\{\mathcal{F}_{t}\}_{t\in\mathbb{T}} ,P)$ be a filtered probability space, with $\mathbb{T}=\mathbb{N}\cup\set{0}$ and $\mathcal{F}_{0}=\{\Omega,\emptyset\}$. For $\cG\subseteq\cF$ we denote by $L^0(\Omega,\cG,P)$, $\widehat{L}^0(\Omega,\cG,P)$ and $\bar{L}^0(\Omega,\cG,P)$ the sets of all $\cG$-measurable random variables with values in $(-\infty,\infty)$, $[-\infty,\infty)$ and $[-\infty,\infty]$, respectively. In addition, we will use the notation $L^{p}:=L^{p}(\Omega,\mathcal{F},P)$, $L^{p}(\cG):=L^{p}(\Omega,\cG,P)$ and $L^{p}_{t}:=L^{p}(\Omega,\mathcal{F}_{t},P)$ for $p\in \set{0,1,\infty}$. Analogous definitions will apply to $\widehat{L}^{0}$ and $\bar{L}^{0}$. Throughout, we will use the convention that $\infty-\infty=-\infty$ and $0\cdot\pm\infty=0$.
In particular we use this convention for $\cF_{t}$-conditional expectation, i.e. for $X\in \bar{L}^{0}$, $E[X|\cF_{t}]:=E[X^{+}|\cF_{t}]-E[X^{-}|\cF_{t}]$, where $X^{+}=(X\vee 0)$ and $X^{-}=(-X\vee 0)$, with $E[X|\cF_{t}]=\lim_{n\to\infty}E[X\wedge n|\cF_{t}]$, for $X\geq 0$.

Throughout this paper, $\cX$ will denote either the space of random variables, i.e. $L^{0}$, or the space of adapted processes, i.e. $\Set{(V_t)_{t\in\bT} \mid V_{t}\in L^{0}_{t}}$. For both cases  we will consider standard pointwise order, understood in the almost sure sense.
In what follows, we will also make use of the multiplication operator denoted as $\cdot_{t}$ and defined by:
\begin{align}
m\cdot_{t}V &:=(V_{0},\ldots,V_{t-1},mV_{t},mV_{t+1},\ldots), \nonumber\\
m\cdot_{t}X &:= mX,\label{eq:conventionV}
\end{align}
for  $V\in\Set{(V_t)_{t\in\bT} \mid V_{t}\in L^{0}_{t}}$, $X\in L^{0}$ and $m\in L^{0}_{t}$.
In order to ease the notation, if no confusion arises, we will drop $\cdot_t$ from the above product, and we will simply write $mV$ and $mX$ instead of $m\cdot_{t}V$ and $m\cdot_{t}X$, respectively.

We note that  $\cX$ is not an $L^0$-module~\cite{FilKupVog2009}, due to the definition of the multiplication~$\cdot_t$. However, in what follows, we will adopt some concepts from $L^0$-module theory, which naturally fit into our study.
Specifically, let $\cK\subseteq\cX$  be an $L^{\infty}_{t}$-convex cone, i.e. $(m_1\cdot_t X+ m_2\cdot_t Y)\in\cK$, for $X,Y\in\cK$ and nonnegative  $m_{1},m_{2}\in L^{\infty}_{t}$.
The map $f:\cK\to\bar{L}^0$ is:
\begin{itemize}
\item {\it $L_{t}^{\infty}$-local} if $\1_{A}f(X)=\1_{A}f(\1_{A}\cdot_t X)$;
\item {\it $L^{\infty}_{t}$-quasi-concave} if $f(\lambda\cdot_t X+(1-\lambda)\cdot_t Y)\geq f(X) \wedge f(Y)$;
\item {\it $L^{\infty}_{t}$-scale invariant} if $f(\beta\cdot_t X)=f(X)$;
\item {\it Monotone} if $X\leq Y \Rightarrow f(X)\leq f(Y)$,
\end{itemize}
for any $X,Y\in\cK,  \ A\in\cF_{t}, \  \lambda,\beta\in L^{\infty}_{t}$ and  $0\leq \lambda\leq1, \beta>0$.

In what follows, we will make use of the sets $\bV$ and  $\widetilde{\bV}$ defined, respectively, by
$$
\bV  := \Set{(V_t)_{t\in\bT} \mid V_t\geq0, \ V_{t}\in L^{0}_{t},  \textrm{ and } V_t=V_{ t\wedge\tau^V }, \  t\in\bT},
$$
and
$$
\widetilde{\mathbb{V}} :=\Set{V\in \mathbb{V} \mid \ V_{t}>0, \ln V_{t}\in L^{1}_t, \ t\in\bT },
$$
where $\tau^V := \inf\set{t\in\bT\mid V_t=0}.$ Note that for any $t\in\bT$, $\bV$ is an $L_{t}^{\infty}$-convex cone contained in $\Set{(V_t)_{t\in\bT} \mid V_{t}\in L^{0}_{t}}$. The elements of $\bV$ can be viewed as (cumulative) value processes of  portfolios of financial securities. In this paper we are primarily interested in portfolios that have integrable growth (cumulative log-return), and this is the reason why we introduced the set $\widetilde{\bV}$.

For $\cK$ equal to $L^{0}$ or equal to $\bV$ we will say that a family $\{f_{t}\}_{t\in\mathbb{T}}$ of mappings $f_{t}:\cK\to\bar{L}^{0}_{t}$ is local, monotone, etc., if for every $t\in\bT$ the mapping $f_{t}$ is $L^{\infty}_{t}$-local, monotone, etc.
Moreover, if $\cK=L^{0}$ then we recall that a family $\{f_{t}\}_{t\in\mathbb{T}}$ of maps $f_{t}: L^{0}\to\bar{L}^0$ is said to be:
\begin{itemize}
\item {\it Cash additive} if for any $t\in\bT$ the function {$f_{t}$ is {\it $L^{0}_{t}$-cash additive}, i.e.} $f_{t}(X+m)=f_{t}(X)+ m$,  for any $X\in L^{0}$ and $m\in L^0_{t}$;
\item {\it Normalized} if $f_{t}(0)=0$, for any $t\in\mathbb{T}$;
\item {\it Strongly time consistent in $L^{1}$} if $f_{s}(X)\geq f_{s}(Y)\Rightarrow f_{t}(X)\geq f_{t}(Y)$, for $s, t\in\mathbb{T}$, $s\geq t$, $X,Y\in L^{1}$.

\item {\it Dynamic risk measure} if $\{-f_{t}\}_{t\in\mathbb{T}}$ is monotone, normalized, cash additive and local;

\item {\it Dynamic certainty equivalent} if there exists  $U:\bar{\mathbb{R}}\rightarrow\bar{\mathbb{R}}$, $U$ strictly increasing and continuous on $\bar{\mathbb{R}}$\footnote{i.e. strictly increasing and continuous of $\bR$, with $U(\pm\infty)=\lim_{n\to\pm\infty}U(n)$.}, such that for any $X\in L^{0}$ and $t\in\mathbb{T}$:
\begin{equation}\label{eq:DCE}
f_{t}(X)=U^{-1}(E[U(X)|\mathcal{F}_{t}]);
\end{equation}
\item {\it Dynamic monetary entropic utility}\footnote{Note that \eqref{eq:entrDRM} is negative of a dynamic entropic risk measure, and by analogy to `dynamic monetary utility function' introduced in \cite{CheDelKup2006}, we use the name `dynamic monetary entropic utility'.} if there exists $\gamma\in\mathbb{R}$, such that for all $t\in\mathbb{T}$, and $X\in L^{0}$,
\begin{equation}\label{eq:entrDRM}
f_{t}(X)=\left\{
\begin{array}{ll}
\frac{1}{\gamma}\ln E[\exp(\gamma X)|\mathcal{F}_{t}] &\quad \textrm{if } \gamma\neq 0\\
E[X|\mathcal{F}_{t}] & \quad \textrm{if } \gamma=0
\end{array}\right.
\end{equation}
In what follows we will denote the dynamic monetary entropic utilities, with parameter $\gamma$, by ${\mu^{\gamma}}_{t\in\bT}$;
\item {\it Dynamic assessment index for random variables} if $\{f_{t}\}_{t\in\mathbb{T}}$ is local, monotone and quasi-concave.
\end{itemize}
On the other hand, if $\cK=\bV$ the family $\{f_{t}\}_{t\in\mathbb{T}}$ of maps $f_{t}: \bV\to\bar{L}^0$ is said to be:
\begin{itemize}
\item {\it Translation invariant} if $f_{t}(X+m\cdot_{t}\1_{\set{k=t}})=f_{t}(X+m\cdot_{t}\1_{\set{s=t}})$ for $t\in\bT$, $m\in L^{0}_{t}$, $X\in\bV$ and $k,s\geq t$, such that $(X+m\cdot_{t}\1_{\set{k=t}})\in\bV$ and $(X+m\cdot_{t}\1_{\set{s=t}})\in\bV$ ;
\item {\it Independent of the past} if $f_{t}(X)=f_{t}(X')$ for $t\in\bT$ and all $X,X'\in\mathbb{V}$ such that $X_{s}=X'_{s}$ for any $s\geq t$;
\item {\it Dynamic assessment index for processes} (DAI) if $\{f_{t}\}_{t\in\mathbb{T}}$ is local, monotone and quasi-concave.
\end{itemize}
Let $f_{t}\colon L^{0}\to \bar{L}^{0}_{t}$, and define a mapping $\widehat{f}_{t}\colon\widehat{L}^0\to \bar{L}^{0}_{t}$ as
\begin{equation}\label{eq:hatf}
\widehat{f}_{t}(X):=\liminf_{n\rightarrow\infty}f_{t}\Big(X\vee (-n)\Big),\quad n\in\bN.
\end{equation}
Clearly, for monotone $f_{t}$, one can replace $\liminf$ with $\lim$ in \eqref{eq:hatf}. Next proposition shows that the function $\widehat{f}_{t}$ inherits  most of the properties of $f_{t}$, although generally speaking, $\widehat{f}_{t}$ is not an extension of $f_{t}$ unless it satisfies the Fatou property\footnote{There are several versions of Fatou property in the existing literature on risk and performance measures, and we use the one from \cite{BiaFri2010}.
We say that $\set{X_{n}}_{n\in\bN}$  is an $L^{0}$-dominated sequence if there exists $Y\in L^{0}$ such that for all $n\in\bN$ we have $|X_{n}|\leq |Y|$.
Function $f$ admitts Fatou property if for any $L^{0}$-dominated sequence such $\set{X_{n}}_{n\in\bN}$  that $X_{n}\xrightarrow{a.s.}X$, we have that $f(X)\geq\limsup_{n\to\infty}f_{t}(X_{n})$.
}
(see Remark~\ref{rem:counterexpleFatouProp}).

\begin{proposition}\label{prop:hatf}
Let $f_{t}\colon L^{0}\to \bar{L}^{0}_{t}$ be $L^{0}_{t}$-local and monotone. Then
\begin{enumerate}[1)]
\item $\widehat{f}_{t}$ is {\it monotone}, i.e.  if $X\geq Y$, then $\widehat{f}_{t}(X)\geq \widehat{f}_{t}(Y)$ for $X,Y\in\widehat{L}^0$;
\item $\widehat{f}_{t}$ is {\it $L^{0}_{t}$-local}, i.e. $\1_{A}\widehat{f}_{t}(X)=\1_{A}\widehat{f}_{t}(\1_{A}X)$ for  $A\in\cF_{t}, X\in\widehat{L}^0$;
\item If $f_{t}$ is $L^{0}_{t}$-cash additive and $f_{t}(0)\neq\infty$, then $\widehat{f}_{t}(X+m)=\widehat{f}_{t}(X)+m$, for $X\in\widehat{L}^0$, $m\in\widehat{L}^{0}_{t}$;
\item $f_{t}(X)=\widehat{f}_{t}(X)$ for $X\in L^{\infty}$. Moreover, if $f_{t}$ has the Fatou property then $f_{t}(X)=\widehat{f}_{t}(X)$ for $X\in L^{0}$.
\end{enumerate}
\end{proposition}
\begin{remark}\label{rem:counterexpleFatouProp}
In general $\widehat{f}_{t}$ might not be an extension of $f_{t}$. For $t=0$ it is sufficient to consider the example $f_{0}(X)=\esssup(X)+\essinf(X)$. This function is monotone and $L^{0}_0$-local.
For $X\sim N(0,1)$ we have $f_{0}(X)=\infty -\infty=-\infty$ and $\widehat{f}_{0}(X)=\liminf_{n\rightarrow\infty}(\infty -n)=\infty$.
\end{remark}

\begin{remark}
{In what follows, the function $f_{t}\colon\widehat{L}_{0}\to \bar{L}^{0}_{t}$ will be understood as $\widehat{f}_{t}$ for corresponding $f_{t}\colon L^{0}\to \bar{L}^{0}_{t}$, i.e. we will drop the notation used in~(\ref{eq:hatf}).}
\end{remark}

\section{Dynamic Limit Growth Indices}\label{S:DLGI}
The main object studied in this paper is the Dynamic Limit Growth Index  and a modification of it that are introduced below.
\begin{definition}
A {\it Dynamic Limit Growth Index} (DLGI) is a family $\{\varphi_{t}\}_{t\in\mathbb{T}}$ of mappings $\varphi_{t}:\mathbb{V}\rightarrow \bar{L}^{0}_{t}$ such that
\begin{equation}\label{eq:DLGI}
\varphi_{t}(V)=\liminf_{T\rightarrow\infty }\frac{\mu_{t}(\ln \frac{V_{T}}{V_{t}})}{T},
\end{equation}
where $\mu_{t}:\widehat{L}^{0}\rightarrow \bar{L}^{0}_{t}$, and $\{\mu_{t}\}_{t\in\mathbb{T}}$ is local and monotone.
We will say that DLGI is {\it risk seeking}, if additionally $\{\mu_{t}\}_{t\in\bT}$ is such that $\mu_{t}(X)=\mu_{t}(X^{+})$ for $t\in\bT$ and $X\in\widehat{L}^{0}$.
\end{definition}

We will often refer to $\{\mu_{t}\}_{t\in\mathbb{T}}$ as a family of mappings that defines DLGI. The maps introduced in Definition~\ref{eq:DLGI} have a natural financial interpretation. The cumulative log-return over the period $(t,T)$ is a common way to measure the process growth. Because it is a random variable, we use a utility measure, say $\mu_{t},$ which represents our preferences (at time $t$). Finally we divide the outcome by $T$ to normalize it in time. Taking the liminf as $T$ goes to infinity allows us to measure the long-time efficiency of our value process.  We use liminf because we want to measure the actual (worst case) efficiency of our portfolio. It also makes this measure more robust (at least to losses).

We also introduce the risk seeking DLGI because it is a more suitable criterion choice for risk seeking investors.
Note that if the family of mappings $\set{\mu_t}_{t\in\bT}$ generates a DLGI, then the family of mappings $\widetilde{\mu}_t(X):=\mu_t(X^+), t\in\bT$, generates a risk seeking DLGI.
A risk seeking DLGI ignores the losses in the sense that it substitutes all losses (negative log returns) by $0$.

We want to use DLGI to assess performance of value processes: the greater the value of DLGI the better the performance of the portfolio.
This is in line with the theory of dynamic assessment indices developed in \cite{BieCiaDraKar2013}.    Therefore, we are interested in identifying conditions under which DLGIs are DAIs. Towards this end, we provide Proposition~\ref{prop:DLGIisAI} that give sufficient and necessary conditions for DLGI to be DAI.

\begin{proposition}\label{prop:DLGIisAI}
Let $\{\varphi_{t}\}_{t\in\mathbb{T}}$ be a DLGI defined  in terms of $\{\mu_{t}\}_{t\in\mathbb{T}}$. Then, $\{\varphi_{t}\}_{t\in\mathbb{T}}$ is DAI if and only if for any $t\in\mathbb{T}$, and any $V\in\mathbb{V}$,
\begin{equation}\label{eq:enough1}
\liminf_{T\rightarrow\infty }\frac{\mu_{t}(\ln \frac{V_{T}}{V_{t}})}{T}=\liminf_{T\rightarrow\infty }\frac{\mu_{t}(\ln V_{T})}{T}.
\end{equation}
\end{proposition}

Relation \eqref{eq:enough1} says that the value of the DLGI at time $t$ is independent of the value of the process $V$ at time $t$.
As mentioned above, the purpose of DLGI is to measure the long term growth of $V$, which intuitively should not depend on the current state.

\begin{remark} An equivalent formulation of condition~(\ref{eq:enough1}) is to require that for any $t\in\bT$, $m\in L^{0}_{t}$ and $\set{X_{T}}_{T\in\bN}$ such that $X_{T}\in \widehat{L}^{0}$ we have that
 $$
 \liminf_{T\rightarrow\infty }\frac{\mu_{t}(X_{T}+m)}{T}=\liminf_{T\rightarrow\infty }\frac{\mu_{t}(X_{T})}{T}.
 $$
In particular, this will be satisfied if there exists a family of maps $f_{t}: L^{0}_{t}\to L^{0}_{t}$ such that for all $X\in \widehat{L}^{0}$,  $|\mu_{t}(X+m)-\mu_{t}(X)|\leq f_{t}(m)$ on the set $\set{\mu_{t}(X)\neq\pm\infty}$, and $\mu_{t}(X+m)=\mu_{t}(X)$ on $\set{\mu_{t}(X)=\pm\infty}$.
For example, if $\mu_t$ is cash-additive then  $f_t(m)=|m|$ (see also Proposition~\ref{prop:DCRMequiv}).
\end{remark}

%$[\ln \1_{A}V_{T}]^{+}=\1_{A}[\ln V_{T}]^{+}$.

\begin{corollary}\label{cor:DGLI-DAI}
Let $\{\mu_{t}\}_{t\in\mathbb{T}}$ be local and monotone, and let $\{\varphi_{t}\}_{t\in\mathbb{T}}$ be a DLGI generated by $\{\mu_{t}\}_{t\in\mathbb{T}}$.
Then $\{\varphi_{t}\}_{t\in\mathbb{T}}$ is adapted, local, scale invariant and independent of the past. Moreover, if  $\{\mu_{t}\}_{t\in\mathbb{T}}$ satisfies \eqref{eq:enough1}, then $\{\varphi_{t}\}_{t\in\mathbb{T}}$ is monotone, quasi-concave and translation invariant.
\end{corollary}
\begin{remark}\label{rem:DLGIisAccI}
Thus, by Corollary~\ref{cor:DGLI-DAI}, any DLGI that is generated by $\set{\mu_t}_{t\in\bT}$ which admits representation~(\ref{eq:enough1}) fulfils all core conditions of Dynamic Acceptability Index introduced in~\cite{BieCiaZha2012} (except of time consistency and positiveness), and for static case introduced in \cite{CheMad2009}.
Recall that Dynamic Acceptability Indices are measures of performance, and hence, DLGI can be seen as a dynamic measure of performance of a given value process.
Similar remark applies  to DLGIs defined as $[\varphi_{t}(V)]^{+}$.
It should be mentioned that this class of maps is not normalized in the sense of~\cite{CheMad2009}\footnote{i.e. $\varphi_{t}(V)=\infty,$ if $V\geq 0$ and $\varphi_{t}(V)=0$, if $V< 0$.}.
\end{remark}

Next we will show that DLGIs that are also DAIs could be easily generated through dynamic risk measures or dynamic certainty equivalents, as shown in the next two propositions.

\begin{proposition}\label{prop:DCRMequiv}
For any dynamic risk measure $\{\rho_{t}\}_{t\in\mathbb{T}}$, the family $\{-\rho_{t}\}_{t\in\mathbb{T}}$ is local, monotone (hence generates a DLGI) and satisfies condition~(\ref{eq:enough1}). Moreover, if $\{\widetilde{\rho}_{t}\}_{t\in\bT}$ is given by $\widetilde{\rho}_{t}(X)=\rho_{t}(X^{+})$, then $\{-\widetilde{\rho}_{t}\}_{t\in\mathbb{T}}$ is also local, monotone and satisfies condition~(\ref{eq:enough1}).
\end{proposition}

\begin{proposition}\label{prop:DCEisDAI}
Let $\{\mu_{t}\}_{t\in\mathbb{T}}$ be a dynamic certainty equivalent. Then,  $\{\mu_{t}\}_{t\in\mathbb{T}}$  is local and monotone (and hence generates a DLGI).
Moreover, if additionally $U$ from \eqref{eq:DCE} is bi-Lipschitz on $\mathbb{R}$ (i.e. $U$ and $U^{-1}$ are Lipschitz\footnote{Although $U$ is defined on $\overline{\bR}$, we require $U$ to be bi-Lipschitz only on $\bR$}.), then $\{\mu_{t}\}_{t\in\mathbb{T}}$ satisfies~\eqref{eq:enough1}.\end{proposition}

\begin{corollary}\label{cor:DRSCisDAI}
By Proposition~\ref{prop:DCRMequiv} and Proposition~\ref{prop:DLGIisAI}, any DLGI generated by $\mu_t=-\rho_t$, $t\in\bT$, with $\set{\rho_t}_{t\in\bT}$ being a dynamic risk measure, is a DAI (for processes).
\end{corollary}

\section{Time consistency of DLGIs}\label{S:time}
One of the key properties in the theory of dynamic risk measures and dynamic performance measures is the dynamic time consistency property.
For risk measures, this property is usually associated with dynamic programming principle (see for instance the review paper~\cite{AccPen2010}), however as shown in~\cite{BieCiaZha2012}  the time consistency for dynamic acceptability indices is of different nature. As we have mentioned in Remark~\ref{rem:DLGIisAccI}, the family of DLGIs is scale-invariant and thus closely related to the latter.
Because of that we introduce the concept of time consistency related to the one introduced in~\cite{BieCiaZha2012}.
As above, $\cX$ will denote the space of random variables $L^{0}$ or adapted processes $\Set{(V_t)_{t\in\bT} \mid V_{t}\in L^{0}_{t}}$, and $\cK\subseteq\cX$.
We are now ready to propose a definition of time consistency.

\begin{definition}\label{41}
We will say that a family $\{f_{t}\}_{t\in\mathbb{T}}$ of maps $f_{t}:\cX\to\bar{L}_{t}^{0}$ is {\it supermartingale time consistent} in $\cK$ if
\begin{equation}\label{eq:timeconsistency}
f_{s}(X)\geq m_{s}\quad \Longrightarrow\quad f_{t}(X)\geq E[m_{s}|\mathcal{F}_{t}],
\end{equation}
for all $s,t\in\bT$ such that $s> t\geq 0$, $X\in\cK$ and $m_{s}\in \bar{L}^{0}_{s}$.
Respectively, we say that $\{f_{t}\}_{t\in\mathbb{T}}$ is \textit{submartingale time consistent} in $\cK$ if
\begin{equation}\label{eq:timeconsistency2}
f_{s}(X)\leq m_{s}\quad \Longrightarrow\quad f_{t}(X)\leq E[m_{s}|\mathcal{F}_{t}],
\end{equation}
for all $s,t\in\bT$ such that $s> t\geq 0$, $X\in\cK$ and $m_{s}\in \bar{L}^{0}_{s}$.
\end{definition}

\begin{remark}
The terminology supermatingale/submartingale time consistent is motivated in Proposition \ref{pr:RejIsSub} below.
\end{remark}

If we only consider $m_s\in \bar{L}^{0}_{t}$ in \eqref{eq:timeconsistency} and in \eqref{eq:timeconsistency2} then we get essentially the definition of time consistency for dynamic acceptability indices introduced in~\cite{BieCiaZha2012}, which shows that our definition is slightly stronger.
Moreover, for $m_s\in \bar{L}^{0}_{t}$ and $\set{f_{t}}_{t\in\bT}$ being a dynamic risk measure the definition of submartingale time consistency \eqref{eq:timeconsistency2} is equivalent to the concept of weak time consistency introduced in~\cite{AccPen2010}.
Hence, our definition of submartingale time consistency is stronger than the definition of weak time consistency (for random variables) studied in~\cite{AccPen2010}.
On the other hand, submartingale time consistency does not imply nor is implied by strong time consistency. For example, the negative of Dynamic Average Value at Risk is submartingale time consistent but not strongly time consistent~\cite{DetSca2005}.  On the contrary, the monetary entropic utility is strongly time consistent but it is not submartingale time consistent for $\gamma>0$, see Proposition~\ref{prop:entropicrisk}.
Analogous reasonings apply with regard to supermartingale time consistency.

The following proposition shows that our definitions of supermartingale/submartingale time consistency can be characterised in terms of supermartingale/submartingale property.

\begin{proposition}\label{pr:RejIsSub}
Let $\{f_{t}\}_{t\in\mathbb{T}}$ be a family of maps $f_{t}:\cX\to\bar{L}_{t}^{0}$. Then
\begin{enumerate}[1)]
\item $\{f_{t}\}_{t\in\mathbb{T}}$ is {\it supermartingale time consistent} in $\cK$ if and only if $\{f_{t}\}_{t\in\mathbb{T}}$ is a supermartingale in $\cK$, i.e. $f_{t}(X)\geq E[f_{s}(X)|\mathcal{F}_{t}]$ for all $X\in\cK$ and $s,t\in\bT$ such that $s>t\geq0$.
\item $\{f_{t}\}_{t\in\mathbb{T}}$ is {\it submartingale time consistent} in $\cK$ if and only if $\{f_{t}\}_{t\in\mathbb{T}}$ is a submartingale in $\cK$, i.e. $f_{t}(X)\leq E[f_{s}(X)|\mathcal{F}_{t}]$ for all $X\in\cK$ and $s,t\in\bT$ such that $s>t\geq0$.
\end{enumerate}
\end{proposition}

We close this section with an intuitive interpretation of our definitions of time consistency. As time evolves, the information about the value process increases in the sense that ${\cal F}_t \subseteq {\cal F}_s,$ for $t\leq s$. Thus, if the index is submartingale time consistent, then one would expect that the additional information will have positive impact on the (conditional) mean value of the index, in the sense that the projection of the future value of the index on the currently available information is no less that the current value of the index. This indeed is confirmed by property 2) in Proposition \ref{pr:RejIsSub}.  On the other hand supermartingale time consistency yields that the impact of the additional information is negative on average. Refer to examples~(\ref{eq:propMart}), (\ref{eq:notRej})~and~(\ref{eq:notAcc}) for more insight.

\section{Dynamic Risk Sensitive Criterion}\label{S:DRSC}
Dynamic analog of Risk Sensitive Criterion~\cite{BiePli1999}, that we study in this section, is one of the most notable examples of DLGI.

\begin{definition}\label{def:DRSC}
A {\it Dynamic Risk Sensitive Criterion} is a family $\{\varphi_{t}^{\gamma}\}_{t\in\mathbb{T}}$ of mappings $\varphi^{\gamma}_{t}:\mathbb{V}\rightarrow \bar{L}^{0}_{t},$ indexed by $\gamma\in\mathbb{R}$, and defined by
\begin{equation}\label{e:DRSC}
\varphi^{\gamma}_{t}(V)=\left\{
\begin{array}{ll}
\liminf_{T\rightarrow\infty }\frac{1}{T}\frac{1}{\gamma}\ln E[V_{T}^{\gamma}|\mathcal{F}_{t}]&\quad \textrm{if $\gamma\neq 0$},\\
\liminf_{T\rightarrow\infty }\frac{1}{T} E[\ln V_{T}|\mathcal{F}_{t}]&\quad \textrm{if $\gamma=0$}.
\end{array}\right.
\end{equation}
\end{definition}

\begin{remark}
It is well known (cf. \cite{DiMSte2007}, and references therein) that for some  processes $V$ that are Markovian, the value of $\varphi^{\gamma}_{t}(V)$ is constant (independent of $t$ in particular). In such cases of course, the analysis carried below trivialises. For example, let $V\in\bV$ be such that $V_{0}>0$ and $V_{t}=V_{0}\exp(\sum_{i=1}^{t}X_{i})$, where $\{X_{t}\}_{t\in\mathbb{T}}$ is adapted, $X_{t}$ is independent of $\mathcal{F}_{t-1}$ and $X_{t}\sim\mathcal N(0,1)$. In this case, $\varphi^{\gamma}_{t}(V)\equiv \frac{\gamma}{2}$. Nevertheless, the class of processes $V$, for which  $\varphi^{\gamma}_{t}(V)$ is a non-constant process, is quite rich; see e.g. (\ref{eq:propMart}) and (\ref{eq:notRej}).
\end{remark}

\noindent We say that the Dynamic Risk Sensitive Criterion is risk-averse if $\gamma<0$, risk neutral if $\gamma=0$, and risk-seeking if $\gamma>0$.  Please note that with $t=0$ we get the standard definition of (static) Risk Sensitive Criterion~\cite{BiePli1999}; in particular, when  $\gamma=0$, the Risk Sensitive Criterion is called the Kelly criterion.

In order to proceed, we first need to recall some facts about Dynamic Monetary Entropic Utilities.

\begin{proposition}\label{prop:entropicrisk}
Let $\{\mu^{\gamma}_{t}\}_{t\in\mathbb{T}}$ be a dynamic monetary entropic utility with $\gamma\in\bR$. Then
\begin{enumerate}[1)]
%\item\label{p:1}  $\{\rho^{\gamma}_{t}\}_{t\in\mathbb{T}}$ is a dynamic risk measure;
\item\label{p:2} $\{\mu^{\gamma}_{t}\}_{t\in\mathbb{T}}$ is a dynamic certainty equivalent;
\item\label{p:3} $\{\mu^{\gamma}_{t}\}_{t\in\mathbb{T}}$ is strongly time consistent in $L^{1}$;
\item\label{p:4} $\{\mu^{\gamma}_{t}\}_{t\in\mathbb{T}}$ is increasing with $\gamma$ in $L^{1}$;
\item\label{p:5} if $\gamma\geq 0$, then $\{\mu^{\gamma}_{t}\}_{t\in\mathbb{T}}$ is supermartingale time consistent in $L^{1}$;
\item\label{p:6}  if $\gamma\leq 0$, $\{\mu^{\gamma}_{t}\}_{t\in\mathbb{T}}$ is submartingale time consistent in $L^{1}$.
\end{enumerate}
\end{proposition}

\noindent For the proof of \ref{p:2}), see e.g.~\cite{KupSch2009}; the proof in~\cite{KupSch2009} is given for the case of $L^{\infty}$, but can be adapted to the case of $L^{0}$. For the proof of~\ref{p:3}), we first need to recall that the dynamic entropic risk measure is upper-semicontinuous in $L^{1}$ (cf.~\cite{BiaFri2010,CheLi2009}), and then  refer to~\cite{BieCiaDraKar2013}. For the proof of~\ref{p:4}), we need to recall that the robust representation of dynamic entropic risk measures holds in the $L^{1}$ framework~\cite{CheLi2009}, and then refer to~\cite{KupSch2009}. Properties \ref{p:5}) and \ref{p:6}) follow directly from property \ref{p:4}), combined with dynamic programming reformulation of property \ref{p:3}); see~\cite{AccPen2010} and \cite[Proposition 6]{DetSca2005}, where the proofs are done for the case of $L^\infty$, but can be adapted to the case of $L^1$.

We are now ready to present the main result of this section. {Arguably, properties 5) and 6) stated in Theorem \ref{th:DRSCprop} are the most interesting ones.}

\begin{theorem}\label{th:DRSCprop}
Let $\gamma\in\mathbb{R}$ and let $\{\varphi_{t}^{\gamma}\}_{t\in\mathbb{T}}$ be a Dynamic Risk Sensitive Criterion. Then
\begin{enumerate}[1)]
\item\label{pp:1} $\{\varphi_{t}^{\gamma}\}_{t\in\mathbb{T}}$  is DLGI generated by $\{\mu^{\gamma}_{t}\}_{t\in\mathbb{T}}$;
\item\label{pp:2} $\{\varphi_{t}^{\gamma}\}_{t\in\mathbb{T}}$  is DAI;
\item\label{pp:3} if $\gamma>0$, then $\big[\varphi^{\gamma}_{t}(V)\big]^{+}$ is a risk seeking DLGI.
\item\label{pp:4} $\{\varphi_{t}^{\gamma}\}_{t\in\mathbb{T}}$ is increasing with $\gamma$ in $\widetilde{\bV}$;
\item\label{pp:5} if $\gamma>0$, then $\{\varphi^{\gamma}_{t}\}_{t\in\mathbb{T}}$ is supermartingale time consistent in $\widetilde{\mathbb{V}}$;
\item\label{pp:6} if $\gamma<0$, then $\{\varphi^{\gamma}_{t}\}_{t\in\mathbb{T}}$ is submartingale time consistent in $\widetilde{\mathbb{V}}$.
\end{enumerate}
\end{theorem}

Next we will show that properties 3), 5) and 6) from Theorem~\ref{th:DRSCprop} are in fact necessary and sufficient conditions {in case of} a large class of filtered probability spaces.

\begin{proposition}\label{pr:DRSCadd}
If $(\Omega,\mathcal{F},\bF,P)$ contains a subspace isomorphic to $([0,1],\mathcal{B}([0,1]),\{\cH\}_{t\in\bT},\lambda)$,  where $\lambda$ is the Borel measure, $\mathcal{H}_{0}$ is trivial, $\mathcal{H}_{1}=\mathcal{B}([0,1])$ and $\cH_{1}=\cH_{2}=\ldots$, then properties 3), 5) and 6) from Theorem~\ref{th:DRSCprop} become if and only if conditions, i.e.
\begin{enumerate}[1)]
\item[3')] if $\gamma\leq 0$, then $\big[\varphi^{\gamma}_{t}(V)\big]^{+}$ is not a risk seeking DLGI;
\item[5')] if $\gamma\leq 0$, then $\{\varphi^{\gamma}_{t}\}_{t\in\mathbb{T}}$ is not supermartingale time consistent in $\widetilde{\mathbb{V}}$;
\item[6')] if $\gamma\geq 0$, then $\{\varphi^{\gamma}_{t}\}_{t\in\mathbb{T}}$ is not submartingale time consistent in $\widetilde{\mathbb{V}}$.
\end{enumerate}
\end{proposition}

\begin{remark}
In particular, Proposition~\ref{pr:DRSCadd} is true for a standard filtered probability space.\footnote{i.e. the spaces which are isomorphic to $([0,1]^{\bN_{0}},\cB([0,1]^{\bN_{0}}),\{\mathcal{F}'_{t}\}_{t\in\bN_{0}},\lambda^{\bN^{0}})$, where $\cB$ is the Borel $\sigma$-algebra, $\lambda^{\bN_{0}}$ is a product of the Borel measures and $\{\mathcal{F}'_{t}\}_{t\in\bN_{0}}$ is the filtration generated by the coordinate functions (cf. \cite{KupSch2009}).}
\end{remark}

We conclude this section by presenting an example that is related to properties~\ref{pp:4},~\ref{pp:5}~and~\ref{pp:6}.

\begin{example}
Let $([0,1],\mathcal{B}([0,1]),\{\mathcal{F}_{t}\}_{t\in\mathbb{N}_{0}},P)$ be a filtered probability space, where $P$ is the standard Lebesgue measure, $\mathcal{F}_{0}$ is trivial and $\mathcal{F}_{t}=\sigma(K^{1}_{t},\ldots,K^{2^t}_{t})$, where $K^{i}_{t}:=[\frac{2(i-1)}{2^{t+1}},\frac{2i}{2^{t+1}}]$. Let $X(\omega)=\omega$ for $\omega\in [0,1],$ and let $\set{\widehat V_{T}}_{T\in\bN_0}$ be defined by
 \begin{equation}\label{eq:propMart}
 \widehat V_{T}(\omega)=e^{T E[X|\mathcal{F}_{T}](\omega)}.
 \end{equation}
We will derive explicit formula for the dynamic risk sensitive criterion $\varphi^{\gamma}_{t}$. We start with the case of $\gamma=-1$. For fixed $t\in\bN_{0}$, we get
$$
\varphi^{-1}_{t}(\widehat V)=\liminf_{T\rightarrow\infty}\frac{-1}{T}\ln E[e^{-T E[X|\mathcal{F}_{T}]}|\mathcal{F}_{t}]=\liminf_{T\rightarrow\infty}(-1)\ln E[(e^{-E[X|\mathcal{F}_{T}]})^{T}|\mathcal{F}_{t}]^{1/T}.
$$
Next for $\omega\in K^{i}_{t}$ and $T\in\bT$, noting that $E[(e^{-E[X|\mathcal{F}_{T}]})^{T}|\mathcal{F}_{t}]^{1/T}(\omega)$ is in fact a power mean, we obtain
\begin{equation}\label{ppp3-1}
\limsup_{T\to\infty}E[(e^{-E[X|\mathcal{F}_{T}]})^{T}|\mathcal{F}_{t}]^{1/T}(\omega)\leq
\limsup_{T\to\infty}[\esssup_{\omega\in K_{t}^{i}}(e^{-E[X|\mathcal{F}_{T}](\omega)})]\leq
\esssup_{\omega\in K_{t}^{i}} e^{-X(\omega)}=e^{-\frac{2(i-1)}{2^{t+1}}}.
\end{equation}
On the other hand using Jensen inequality, for any $T_{0}\in\bT$, such that $T_{0}>t$, we get
\begin{align}
\limsup_{T\to\infty}E[(e^{-E[X|\mathcal{F}_{T}]})^{T}|\mathcal{F}_{t}]^{1/T}(\omega) & =\limsup_{T\to\infty}E[E[e^{-TE[X|\cF_{T}]}|\cF_{T_{0}}]|\cF_{t}]^{1/T}(\omega)\notag \\
& \geq \limsup_{T\to\infty}E[e^{-T E[E[X|\cF_{T}]|\mathcal{F}_{T_{0}}]}|\cF_{t}]^{1/T}(\omega)\notag \\
& =\limsup_{T\to\infty}E[(e^{-E[X|\mathcal{F}_{T_{0}}]})^{T}|\cF_{t}]^{1/T}(\omega)\notag \\
& =\esssup_{\omega\in K_{t}^{i}} e^{-E[X|\cF_{T_{0}}]}=e^{-(\frac{2(i-1)}{2^{t+1}}+\frac{1}{2^{T_{0}+1}})}.\label{ppp3-2}
\end{align}
Letting $T_{0}\to\infty$, and combining \eqref{ppp3-1} with \eqref{ppp3-2}, we conclude that for $\omega\in K^{i}_{t}$,
$$
\varphi^{-1}_{t}(\widehat V)(\omega)=(-1)\ln e^{-\frac{2(i-1)}{2^{t+1}}}=\frac{2(i-1)}{2^{t+1}}.
$$
Using similar computations, it is easy to show that, for $\gamma\in\bR$ and $\omega\in K^{i}_{t}$, we have
$$
\varphi^{\gamma}_{t}(\widehat V)(\omega)=\left\{
\begin{array}{ll}
\frac{2(i-1)}{2^{t+1}} & \gamma<0,\\
\frac{2(i-1)+2i}{2^{t+2}} & \gamma=0,\\
\frac{2i}{2^{t+1}} & \gamma>0.
\end{array}\right.
$$
Now, it clear from the above formula that $\varphi^{\gamma}_{t}(\widehat V)$ is increasing in $\gamma,$ so that property 4) is fulfilled. In addition, one can easily check that process $\varphi^{\gamma}_{t}(\widehat V)$ is a submartingale (resp. supermartingale), with respect to the filtration $\{\mathcal{F}_{t}\}_{t\in\mathbb{N}_{0}}$, when $\gamma<0$ (resp. $\gamma>0$).

It is interesting to note that the values of $\varphi^{\gamma}_{t}(\widehat V)$ are separated into three regimes: risk-seeking ($\gamma >0$), risk-neutral ($\gamma =0$) and risk-averse ($\gamma <0$).
\end{example}

\begin{appendix}

\section{Appendix}\label{S:App}
\begin{proof}[\textbf{Proof of Proposition~\ref{prop:hatf}.}]
Let $f_{t}:L^{0}\to \bar{L}^{0}_{t}$ be local and monotone.

\smallskip
\noindent 1) Monotonicity follows immediately.

\smallskip
\noindent 2) As for locality,  we have
\begin{align*}
\1_{A}\widehat{f}_{t}(\1_{A}X) & =\1_{A}\lim_{n\rightarrow\infty}f_{t}\Big((\1_{A}X)\vee (-n)\Big)=\1_{A}\lim_{n\rightarrow\infty}f_{t}\Big(\1_{A}(X\vee (-n))\Big)\\
 & =\lim_{n\rightarrow\infty}\1_{A} f_{t}\Big(\1_{A}(X\vee (-n))\Big)=\lim_{n\rightarrow\infty}\1_{A} f_{t}\Big(X\vee (-n)\Big)
 =\1_{A}\lim_{n\rightarrow\infty} f_{t}\Big(X\vee (-n)\Big)\\
 &=\1_{A}\widehat{f}_{t}(X),
\end{align*}
where we use appropriately the convention $0\cdot\infty=0$.

\smallskip
\noindent 3) Assume that $f_{t}$ is cash additive and let $X\in\widehat{L}^0(\Omega,\cF,P)$. First, we will prove cash additivity of $\widehat{f}_{t}$ for $m\in L^{0}_{t}$.
We know that
\begin{align*}
\widehat{f}_{t}(X+m) & =\lim_{n\rightarrow\infty}f_{t}\Big((X+m)\vee (-n)\Big)=\lim_{n\rightarrow\infty}f_{t}\Big(X\vee (-n-m)+m\Big)\\
& =\lim_{n\rightarrow\infty}f_{t}\Big(X\vee (-n-m)\Big)+m.
\end{align*}
Thus, it is enough to show that
\begin{equation}\label{eq:prop1a1}
\widehat{f}_{t}(X)=\lim_{n\rightarrow\infty}f_{t}\Big(X\vee (-n-m)\Big).
\end{equation}
For any $k\in\bN$, we have that
$$
\1_{\set{-k<m<k}} \Big[X\vee (-n-k)\Big]\leq \1_{\set{-k<m<k}} \Big[X\vee (-n-m)\Big] \leq \1_{\set{-k<m<k}} \Big[X\vee (-n+k)\Big].
$$
Due to $L^{\infty}_{t}$-locality of $f_{t}$, we get that
$$
\1_{\set{-k<m<k}}\widehat{f}_{t}(X)=\1_{\set{-k<m<k}}\lim_{n\rightarrow\infty}f_{t}\Big(X\vee (-n-m)\Big).
$$
Since $m\in L^{0}_{t}$, we have that $P[\set{-k<m<k}]\to 1$ as $k\to\infty$ which proves the equality \eqref{eq:prop1a1}.

Now, let $m\in\widehat{L}^0(\Omega,\cF,P)$. Using the above result, and because of locality of $\hat f_t$ and the fact that $\1_{\set{m>-\infty}}m\in L^{0}_{t}$, we deduce that
$$
\1_{\set{m>-\infty}}\widehat{f}_{t}(X+m)=\1_{\set{m>-\infty}}(\widehat{f}_{t}(X)+m).
$$
On the other hand
\begin{align*}
\1_{\set{m=-\infty}}\widehat{f}_{t}(X+m) & =\1_{\set{m=-\infty}}\lim_{n\rightarrow\infty}f_{t}((-\infty)\vee (-n))=\1_{\set{m=-\infty}}\lim_{n\rightarrow\infty}(f_{t}(0)-n)\\
& =\1_{\set{m=-\infty}}(-\infty)=\1_{\set{m=-\infty}}(\widehat{f}_{t}(X)+m).
\end{align*}
Combining those above two equalities, cash-additivity of $\hat{f}_t$ follows immediately.

\smallskip
\noindent 4) If $X\in L^{\infty}$, then there exists $n\in\bN$ such that $X\vee(-n)=X$  which concludes the proof. Now let $X\in L^{0}$ and let us assume that $f_{t}$ has the Fatou property. Put $X_{n}:=X\vee(-n)$ for $n\in\bN$. The sequence $\{X\}_{n\in\bN}$ is $L^{0}$- dominated by $X$. Moreover $X_{n}\xrightarrow{a.s.} X$. Hence, we have that
$$
\widehat{f}_{t}(X)=\lim_{n\to\infty}f_{t}(X_{n})\leq \limsup_{n\to\infty}f_{t}(X_{n})\leq f_{t}(X)\leq \lim_{n\to\infty} f_{t}(X_{n})=\widehat{f}_{t}(X),
$$
where the last inequality is the consequence of the fact that for any $n\in\bN$ we have $X\leq X_{n}$, which implies $f_{t}(X)\leq f_{t}(X_{n})$.
\end{proof}

\begin{proof}[\textbf{Proof of Proposition~\ref{prop:DLGIisAI}.}]
Let $\{\varphi_{t}\}_{t\in\mathbb{T}}$ be DLGI generated by $\{\mu_{t}\}_{t\in\mathbb{T}}$, and thus $\{\mu_{t}\}_{t\in\mathbb{T}}$ is local and monotone.

\smallskip
\noindent ($\Leftarrow$) Let $\{\mu_{t}\}_{t\in\mathbb{T}}$ satisfy~(\ref{eq:enough1}), and we will show that $\{\varphi_{t}\}_{t\in\mathbb{T}}$ is a DAI.

Monotonicity is straightforward. Let $V,V'\in\mathbb{V}$, such that $V\geq V'$. We will show that $\varphi_{t}(V)\geq\varphi_{t}(V')$ for any $t\in\mathbb{T}$.
 Consider $t,T\in\mathbb{T}$, such that $T\geq t$. Since $V_{T}\geq V'_{T}$, we have that $\ln V_{T}\geq \ln V'_{T}$, and consequently $\frac{\mu_{t}(\ln V_{T})}{T}\geq \frac{\mu_{t}(\ln V'_{T})}{T}$, for any $T\geq t$. Hence,
$$
\liminf_{T\rightarrow\infty}\frac{\mu_{t}(\ln V_{T})}{T}\geq \liminf_{T\rightarrow\infty}\frac{\mu_{t}(\ln V'_{T})}{T}.
$$

Next we prove locality. Let us fix $t\in\mathbb{T}$ and $A\in\mathcal{F}_{t}$. For $T\geq t$, using locality of $\mu_{t}$ and the convention $0\cdot\infty=0$, we deduce
\begin{align*}
\mathbf{1}_{A}\varphi_{t}(\mathbf{1}_{A}\cdot_{t}V) & =\mathbf{1}_{A}\liminf_{T\rightarrow\infty }\frac{\mu_{t}(\ln \mathbf{1}_{A}V_{T})}{T}=\liminf_{T\rightarrow\infty }\frac{\mathbf{1}_{A}\mu_{t}(\ln \mathbf{1}_{A}V_{T})}{T}\\
& =\liminf_{T\rightarrow\infty }\frac{\mathbf{1}_{A}\mu_{t}(\mathbf{1}_{A}\ln \mathbf{1}_{A}V_{T})}{T}=\liminf_{T\rightarrow\infty }\frac{\mathbf{1}_{A}\mu_{t}(\mathbf{1}_{A}\ln V_{T}+\mathbf{1}_{A}\ln \mathbf{1}_{A})}{T}\\
& =\liminf_{T\rightarrow\infty }\frac{\mathbf{1}_{A}\mu_{t}(\mathbf{1}_{A}\ln V_{T})}{T}=\liminf_{T\rightarrow\infty }\frac{\mathbf{1}_{A}\mu_{t}(\ln V_{T})}{T}=\mathbf{1}_{A}\varphi_{t}(V).
\end{align*}

Finally, let us prove quasiconcavity. Let $t\in\mathbb{T}$, $V, V'\in \bV$ and $\lambda\in L^{0}_{t}$, $0\leq \lambda\leq 1$. Without loss of generality, due to locality of  $\mu_{t}$, we assume that $0<\lambda<1$. Since log is monotone,  and $V,V'\geq0$, we have
\begin{align*}
\varphi_{t}(\lambda\cdot_{t}V+(1-\lambda)\cdot_{t}V') & =\liminf_{T\rightarrow\infty }\frac{\mu_{t}(\ln [\lambda V_{T}+(1-\lambda)V'_{T}])}{T} \\
& \geq \liminf_{T\rightarrow\infty }\Big[\min{\Big\{\frac{\mu_{t}(\ln \lambda V_{T})}{T},\frac{\mu_{t}(\ln (1-\lambda)V'_{T})}{T}\Big\}}\Big]\\
& =\min{\Big(\liminf_{T\rightarrow\infty }\frac{\mu_{t}(\ln V_{T})+\ln\lambda}{T},\liminf_{T\rightarrow\infty }\frac{\mu_{t}(\ln V'_{T})+\ln(1-\lambda)}{T}\Big)}\\
& = \varphi_{t}(V)\wedge\varphi_{t}(V'),
\end{align*}
which completes this part of the proof.

\smallskip
\noindent ($\Rightarrow$) Assume that $\{\varphi_{t}\}_{t\in\mathbb{T}}$ is a DAI. Let $t\in\mathbb{T}$, $V\in\mathbb{V}$, and define $V'_{s}=V_{s}$ for $s\neq t$, and $V'_{t}=\min{(1,V_{t})}$. Note that $V'\in\mathbb{V}$, and $V\geq V'$. By monotonicity of $\varphi_{t}$, we get
$$
\liminf_{T\rightarrow\infty}\frac{\mu_{t}(\ln \frac{V_{T}}{V_{t}})}{T}\geq \liminf_{T\rightarrow\infty}\frac{\mu_{t}(\ln \frac{V'_{T}}{V'_{t}})}{T},
$$
and due to $L^{0}_{t}$-locality of $\mu_{t}$,  we continue
$$
1_{\{V_{t}\geq 1\}}\liminf_{T\rightarrow\infty}\frac{\mu_{t}(\ln \frac{V_{T}}{V_{t}})}{T}\geq 1_{\{V_{t}\geq 1\}}\liminf_{T\rightarrow\infty}\frac{\mu_{t}(1_{\{V_{t}\geq 1\}}\ln \frac{V'_{T}}{V'_{t}})}{T}.
$$
Next, since $V'_t=1$ on the set $\{V_{t}\geq 1\}$, we have
$$
1_{\{V_{t}\geq 1\}}\liminf_{T\rightarrow\infty}\frac{\mu_{t}(\ln \frac{V_{T}}{V_{t}})}{T}\geq 1_{\{V_{t}\geq 1\}}\liminf_{T\rightarrow\infty}\frac{\mu_{t}(1_{\{V_{t}\geq 1\}}\ln V'_{T})}{T},
$$
and since $V_{T}=V'_{T}$ for $T>t$, we finally conclude
$$
1_{\{V_{t}\geq 1\}}\liminf_{T\rightarrow\infty}\frac{\mu_{t}(\ln \frac{V_{T}}{V_{t}})}{T}\geq 1_{\{V_{t}\geq 1\}}\liminf_{T\rightarrow\infty}\frac{\mu_{t}(\ln V_{T})}{T}.
$$
Note that $1_{\{V_{t}\geq 1\}}\ln\frac{V_{T}}{V_{t}}\leq 1_{\{V_{t}\geq 1\}}\ln V_{T}$ for $T>t$.
By monotonicity of $\mu_{t}$, we get
$$
1_{\{V_{t}\geq 1\}}\liminf_{T\rightarrow\infty}\frac{\mu_{t}(\ln \frac{V_{T}}{V_{t}})}{T}\leq 1_{\{V_{t}\geq 1\}}\liminf_{T\rightarrow\infty}\frac{\mu_{t}(\ln V_{T})}{T}.
$$
Combining the above inequalities, we have that equality \eqref{eq:enough1} holds true on set ${\{V_{t}\geq1\}}$. The proof for the set ${\{V_{t}<1\}}$ is similar.
\end{proof}

\begin{proof}[\textbf{Proof of Proposition~\ref{prop:DCRMequiv}.}]
Let $\{\rho_{t}\}_{t\in\mathbb{T}}$ be a dynamic risk measure. Monotonicity and locality of  $\{-\rho_{t}\}_{t\in\mathbb{T}}$ follow directly from the definition of dynamic risk measures.
Let us fix $t\in\mathbb{T}$.
First we will prove that condition~(\ref{eq:enough1}) is satisfied by $\{-\rho_{t}\}_{t\in\mathbb{T}}$.
For $V\in\mathbb{V}$, we have
$$
\liminf_{T\rightarrow\infty }\frac{-\rho_{t}(\ln \frac{V_{T}}{V_{t}})}{T}=\liminf_{T\rightarrow\infty }\frac{-\rho_{t}(\ln V_{T})-\ln V_{t}}{T}=\liminf_{T\rightarrow\infty }\frac{-\rho_{t}(\ln V_{T})}{T}.
$$
The above equality is straightforward on set $\{V_{t}>0\}$, since $\frac{\ln V_{t}}{T}\rightarrow 0$, $T\rightarrow \infty$.
On the set $\{V_{t}=0\}$,  we have that $\1_{\set{V_{t}=0}}V_{T}= 0$, and by locality and normalization of $-\rho_{t}$,  we get that both sides are equal to $(-\infty)$.

Next, monotonicity and locality of  $\{-\widetilde{\rho_t}\}_{t\in\bT}$ is straightforward. We will show now that \eqref{eq:enough1} also holds true for $\{\widetilde{\rho_t}\}_{t\in\bT}$.
Let $V\in\mathbb{V}$. On the $\mathcal{F}_{t}$-measurable set $\{V_{t}=0\}$ both sides of \eqref{eq:enough1} are equal to 0.
Due to this, and locality of $\widetilde{\rho}_t$, we can assume that $P[V_{t}>0]=1$.
Then, it is easy to note that
\begin{align*}
\liminf_{T\rightarrow\infty }\frac{-\rho_{t}([\ln \frac{V_{T}}{V_{t}}]^{+})}{T} & =\liminf_{T\rightarrow\infty }\frac{-\rho_{t}(\1_{\{V_{T}>V_{t}\}}\ln \frac{V_{T}}{V_{t}})}{T}\\
& =\liminf_{T\rightarrow\infty }\frac{-\rho_{t}(\1_{\{V_{T}>V_{t}\}}\ln V_{T}-\1_{\{V_{T}>V_{t}\}}\ln V_{t})}{T}.
\end{align*}
Also, one can easily deduce the following inequalities
$$
\1_{\{V_{T}>1\}}\ln{V_{T}}-2|\ln{V_{t}}| \leq \1_{\{V_{T}>V_{t}\}}\ln V_{T}-\1_{\{V_{T}>V_{t}\}}\ln V_{t} \leq \1_{\{V_{T}>1\}}\ln{V_{T}}+|\ln{V_{t}}|.
$$
From the above, and monotonicity of the dynamic risk measure,  we get
$$
\liminf_{T\rightarrow\infty }\frac{-\rho_{t}([\ln V_{T}]^{+}-2|\ln V_{t}| )}{T}
\leq
\liminf_{T\rightarrow\infty }\frac{-\rho_{t}([\ln \frac{V_{T}}{V_{t}}]^{+})}{T}
\leq
\liminf_{T\rightarrow\infty }\frac{-\rho_{t}([\ln V_{T}]^{+}+2|\ln V_{t}| )}{T}.
$$
Since $-\rho_{t}$ is cash additive, continue
$$
\liminf_{T\rightarrow\infty }\frac{-\rho_{t}([\ln V_{T}]^{+}\pm 2|\ln V_{t}| )}{T}=\liminf_{T\rightarrow\infty }\frac{-\rho_{t}([\ln V_{T}]^{+})\pm 2|\ln V_{t}|}{T}=\liminf_{T\rightarrow\infty }\frac{-\rho_{t}([\ln V_{T}]^{+})}{T},
$$
which concludes the proof.
\end{proof}

\begin{proof}[\textbf{Proof of Proposition~\ref{prop:DCEisDAI}}]
Let $\{\mu_{t}\}_{t\in\mathbb{T}}$ be a dynamic certainty equivalent defined as in \eqref{eq:DCE}, with $U$ being a continuous an increasing function.
Clearly $\mu_{t}$ is $\mathcal{F}_{t}$-measurable.

Monotonicity is straightforward. Let us fix $t\in\mathbb{T}$. Let $X,Y\in \widehat{L}^{0}$, $X\geq Y$. Because $U$ is increasing transform we get $U(X)\geq U(Y)$, and $E[U(X)|\mathcal{F}_{t}]\geq E[U(Y)|\mathcal{F}_{t}]$. Now, $U^{-1}$ is also an increasing function, so $U^{-1}(E[U(X)|\mathcal{F}_{t}])\geq U^{-1}(E[U(Y)|\mathcal{F}_{t}])$.

Next we prove locality. Note that any deterministic function, in particular $U$ and $U^{-1}$, is local.
Thus, for any $t\in\mathbb{T}$ and $A\in\mathcal{F}_{t}$, we have
\begin{align*}
\1_A\mu_t(X) &= \1_{A}U^{-1}(E[U(X)|\mathcal{F}_{t}]) = \1_{A}U^{-1}(\1_AE[U(X)|\mathcal{F}_{t}]) \\
& = \1_{A}U^{-1}(E[1_A U(X)|\mathcal{F}_{t}]) = \1_{A}U^{-1}(E[U(1_AX)|\mathcal{F}_{t}]) \\
& = \1_A\mu_t(1_AX),
\end{align*}
which proves locality of $\mu_t$.

Finally  we will prove the second part of the Proposition~\ref{prop:DCEisDAI}.
Let $U$ be a bi-Lipschitz function with $L_{U}\in\mathbb{R}$ and $L_{U^{-1}}\in\mathbb{R}$ being the corresponding Lipschitz constants.
Consider $t\in\mathbb{T}$ and $V\in\mathbb{V}$.
On $\cF_t$-measurable set $\set{V_t=0}$, $\1_{\{V_{t}=0\}}V_{T}=0$, and hence both sides of \eqref{eq:enough1} are equal to $-\infty$.

From now on we make a (reasonable) assumption that $P[V_{t}>0]>0$, which due to locality of $\mu_t$, allows us to assume that $P[V_{t}>0]=1$.

First we prove that for a fixed  $T\in\bT$, we get
\begin{equation}\label{eq1U1}
\{U^{-1}(E[U(\ln V_{T})|\mathcal{F}_{t}])=-\infty\} =\{U^{-1}(E[U(\ln \frac{V_{T}}{V_{t}})|\mathcal{F}_{t}])=-\infty\}.
\end{equation}
As $U$ is strictly increasing we know that~\eqref{eq1U1} is equivalent to
\begin{equation}\label{eq1U1.1}
\{E[U(\ln V_{T})|\mathcal{F}_{t}]=U(-\infty)\} =\{E[U(\ln \frac{V_{T}}{V_{t}})|\mathcal{F}_{t}]=U(-\infty)\}.
\end{equation}
Next we consider two cases: a) $U(-\infty)>-\infty$ and b) $U(-\infty)=-\infty$.

\smallskip
\noindent Case a) It is clear that the set $\{E[1_{\set{V_{T}=0}}|\cF_{t}]=1\}$ is the subset of both sets in \eqref{eq1U1.1}. Thus, it is sufficient to show that
\begin{equation}\label{eq1Uc1}
P\Big[\{E[U(\ln V_{T})|\mathcal{F}_{t}]=U(-\infty)\} \cap \{E[1_{\set{V_{T}>0}}|\cF_{t}]>0\}\Big]=0
\end{equation}
and
\begin{equation}\label{eq1Uc2}
P\Big[\{E[U(\ln \frac{V_{T}}{V_{t}})|\mathcal{F}_{t}]=U(-\infty)\} \cap \{E[1_{\set{V_{T}>0}}|\cF_{t}]>0\}\Big]=0.
\end{equation}
Let us prove \eqref{eq1Uc1}. Let
$$
B:=\{E[U(\ln V_{T})|\mathcal{F}_{t}]=U(-\infty)\} \cap \{E[1_{\set{V_{T}>0}}|\cF_{t}]>0\}.
$$
Note that $B\in\cF_{t}$.  On the contrary let us assume that $P[B]>0$. Then
$$
P[\set{V_{T}>0}\cap B]= E[1_{B}E[1_{\set{V_{T}>0}}|\cF_{t}]]>0.
$$
Because ${\set{V_{T}>0}}\cap B=\bigcup_{n\in\bN} {\set{V_{T}>\frac{1}{n}}}\cap B$, we know that there exists $n_{0}\in\bN$, such that $P[\set{V_{T}>\frac{1}{n_{0}}}\cap B]>0$.
Using that we obtain
\begin{align}
E[1_{B}E[U(\ln V_{T})|\mathcal{F}_{t}]] &=E[1_{B}E[1_{\set{V_{T}>\frac{1}{n_{0}}}}U(\ln V_{T})+1_{\set{V_{T}\leq\frac{1}{n_{0}}}}U(\ln V_{T})|\cF_{t}]]\notag\\
& \geq E[1_{B}E[1_{\set{V_{T}>\frac{1}{n_{0}}}}U(\ln \frac{1}{n_{0}})+1_{\set{V_{T}\leq\frac{1}{n_{0}}}}U(-\infty)|\cF_{t}]]\notag\\
& =E[1_{B\cap\set{V_{T}>\frac{1}{n_{0}}}}U(\ln \frac{1}{n_{0}})+1_{B\cap\set{V_{T}\leq\frac{1}{n_{0}}}}U(-\infty)]\notag\\
& >E[1_{B}U(-\infty)].\label{eq1Uc3}
\end{align}
Inequality \eqref{eq1Uc3} jointly with the definition of $B$ leads to contradiction with the assumption that $P(B)>0$, which verifies that \eqref{eq1Uc1} is true. The proof of \eqref{eq1Uc2} is analogous, since $P(V_t>0)=1$.

\smallskip
\noindent Case b) It is enough to show that
\begin{equation}\label{eq1Ucc1}
\{E[U(\ln V_{T})|\mathcal{F}_{t}]=-\infty\}=\{E[U(\ln \frac{V_{T}}{V_{t}})|\mathcal{F}_{t}]=-\infty\}.
\end{equation}
Now, because $U$ is Lipschitz and $V_{t}>0$, then, on the set  $\set{V_{T}>0}$  we get
\begin{equation}\label{eq1U2}
U(\ln V_{T})-L_{U}|\ln V_{t}|\leq U(\ln \frac{V_{T}}{V_{t}})\leq U(\ln V_{T})+L_{U}|\ln V_{t}|.
\end{equation}
In addition, the above inequalities obviously hold true on the set $\set{V_{T}=0}$, as on this set we have $U(\ln V_{T})=U(\ln\frac{V_{T}}{V_{t}})=U(-\infty)=-\infty$. Consequently,
\begin{equation}\label{eq1U2a}
E[U(\ln V_{T})|\cF_{t}]-L_{U}|\ln V_{t}|\leq E[U(\ln \frac{V_{T}}{V_{t}})|\cF_{t}]\leq E[U(\ln V_{T})|\cF_{t}]+L_{U}|\ln V_{t}|.
\end{equation}
Analogously, we obtain
\begin{equation}\label{eq1U3}
E[U(\ln \frac{V_{T}}{V_{t}})|\cF_{t}]-L_{U}|\ln V_{t}|\leq E[U(\ln V_{T})|\cF_{t}]\leq E[U(\ln \frac{V_{T}}{V_{t}})|\cF_{t}]+L_{U}|\ln V_{t}|.
\end{equation}
Combining \eqref{eq1U2a} and \eqref{eq1U3}, we obtain equality \eqref{eq1Ucc1}. So, \eqref{eq1U1} has been demonstrated.

Next, noting that $V_{T}<\infty$, and applying similar reasoning as in the proof of \eqref{eq1U1}, one can show that
\begin{equation}\label{eq1U4}
\{U^{-1}(E[U(\ln V_{T})|\mathcal{F}_{t}])=+\infty\} =\{U^{-1}(E[U(\ln \frac{V_{T}}{V_{t}})|\mathcal{F}_{t}])=+\infty\}.
\end{equation}
Now, let
$$
K^{-}_{T}:=\{U^{-1}(E[U(\ln V_{T})|\mathcal{F}_{t}])=-\infty\},\quad K^{+}_{T}:=\{U^{-1}(E[U(\ln V_{T})|\mathcal{F}_{t}])=\infty\},\quad T\in \bT.
$$
Combining \eqref{eq1U1} and \eqref{eq1U4} we obtain $\mu_t(\ln V_{T})=\mu_{t}(\ln\frac{V_{T}}{V_{t}})$, on $\cF_{t}$-measurable set $K^{-}_{T}\cup K^{+}_{T}$. On the set $(K^{-}_{T}\cup K^{+}_{T})^{c}$ we get $|\mu_t(\ln V_{T})|<\infty$ and $|\mu_t(\ln \frac{V_{T}}{V_{t}})|<\infty$. Moreover, since $U$ is strictly increasing we also get $|E[U(\ln V_{T})|\cF_{t}]|<\infty$ and $|E[U(\ln \frac{V_{T}}{V_{t}})|\cF_{t}]|<\infty$.
Thus, using the fact that $U$ is bi-Lipschitz, then, on set $(K^{-}_{T}\cup K^{+}_{T})^{c}$, we get
\begin{align}
|U^{-1}(E[U(\ln \frac{V_{T}}{V_{t}})|\mathcal{F}_{t}])-U^{-1}(E[U(\ln V_{T})|\mathcal{F}_{t}])| &\leq L_{U^{-1}}|E[U(\ln \frac{V_{T}}{V_{t}})|\mathcal{F}_{t}]-E[U(\ln V_{T})|\mathcal{F}_{t}]|\notag \\
& \leq L_{U^{-1}}L_{U}|\ln V_{t}|.\label{eq1U5}
\end{align}
We are now finally ready to prove the main statement. Let
$$
K^{-}:=\{\omega\in\Omega: \sum_{T\in\bT}1_{K^{-}_{T}}(\omega)<\infty\},\quad K^{+}:=\{\omega\in\Omega: \sum_{T\in\bT}1_{(K^{+}_{T})^{c}}(\omega)=\infty\}.
$$
Using \eqref{eq1U5}, on the set $K^{-}\cap K^{+}$ we obtain
$$
\liminf_{T\rightarrow\infty }\frac{|\mu_{t}(\ln \frac{V_{T}}{V_{t}})-\mu_{t}(\ln V_{T})|}{T}\leq\liminf_{T\rightarrow\infty }\frac{L_{U}L_{U^{-1}}|\ln V_{t}|}{T}=0.
$$
which proves the equality \eqref{eq:enough1} on this set. Using \eqref{eq1U1}  we get the equality \eqref{eq:enough1} on $(K^{-})^{c}$; similarly, using \eqref{eq1U4} we get  \eqref{eq:enough1} on $(K^{+})^{c}$. This completes the proof.
\end{proof}

\begin{proof}[\textbf{Proof of Proposition~\ref{pr:RejIsSub}}]
We will prove only the supermartingale part (proof for submartingale is similar).

\smallskip
\noindent $(\Rightarrow)$ Let $m_{s}=f_{s}(V)$. Because $f_{s}(V)\geq f_{s}(V)$, using \eqref{eq:timeconsistency}, we get $f_{t}(V)\geq E[f_{s}(V)|\mathcal{F}_{t}]$.

\smallskip
\noindent $(\Leftarrow)$ Let $m_{s}$ be such that $f_{s}(V)\geq m_{s}$. Using this, and the fact that $f_t(V)$ is supermartingale, we immediately get
$$
f_{t}(V)\geq E[f_{s}(V)|\mathcal{F}_{t}] \geq E[m_{s}|\mathcal{F}_{t}].
$$
This concludes the proof.
\end{proof}

\begin{proof}[\textbf{Proof of Theorem~\ref{th:DRSCprop}}]
For a fixed $\gamma\in\mathbb{R}$, let $\{\varphi_{t}^{\gamma}\}_{t\in\mathbb{T}}$ be a Dynamic Risk Sensitive Criterion.

\smallskip
\noindent 1) It is enough to show that
\begin{equation}\label{eq:prop4.4-1}
\varphi^{\gamma}_{t}(V)=\liminf_{T\rightarrow\infty }\frac{\mu^{\gamma}_{t}(\ln \frac{V_{T}}{V_{t}})}{T}, \quad t\in\bT, \ V\in\bV.
\end{equation}
Note that on $\cF_t$-measurable set $\set{V_t=0}$, $\1_{\{V_{t}=0\}}V_{T}=0$, and hence both sides of \eqref{eq:prop4.4-1} are equal to $-\infty$.
Thus, due to locality of $\mu_t$, it is enough to consider the case $P[V_{t}>0]=1$.

For fixed $V\in\mathbb{V}$ and $t\in\mathbb{T}$ we have
$$
\liminf_{T\rightarrow\infty }\frac{\mu_{t}^{\gamma}(\ln\frac{V_{T}}{V_{t}})}{T}
=\liminf_{T\rightarrow\infty }\frac{\ln E[\exp(\gamma \ln \frac{V_{T}}{V_{t}})|\mathcal{F}_{t}]}{\gamma T}
=\liminf_{T\rightarrow\infty }\Big[\frac{1}{T}\frac{1}{\gamma}\ln E[V_{T}^{\gamma}|\mathcal{F}_{t}]-\frac{1}{T}\ln V_{t}\Big]=\varphi_{t}^{\gamma}(V).
$$
For $\gamma=0$, we immediately get
$$
\liminf_{T\rightarrow\infty }\frac{\mu^{0}_{t}(\ln\frac{V_{T}}{V_{t}})}{T}=\liminf_{T\rightarrow\infty}\Big[\frac{E[\ln V_{T}|\mathcal{F}_{t}]}{T}-\frac{\ln V_{t}}{T}\Big]=\liminf_{T\rightarrow\infty}\frac{1}{T}E[\ln V_{T}|\mathcal{F}_{t}] = \varphi^{0}_{t}(V).
$$

\smallskip
\noindent 2) It is an immediate result of Corollary~\ref{cor:DRSCisDAI} and 1), since $\{-\mu^{\gamma}_{t}\}_{t\in\mathbb{T}}$ is a dynamic risk measure.

\smallskip
\noindent 3) It is enough to show that for $\gamma>0$ we have
\begin{equation}\label{eq:DRSCisMLGI}
\big[\varphi^{\gamma}_{t}(V)\big]^{+}=\liminf_{T\rightarrow\infty }\frac{\mu^{\gamma}_{t}([\ln\frac{V_{T}}{V_{t}}]^{+})}{T}.
\end{equation}
As in the previous case, without loss of generality, we can assume that $P[V_{t}>0]=1$.
For every $t\in\mathbb{T}$ and $V\in\mathbb{V}$, we deduce

\begin{align}
\liminf_{T\rightarrow\infty }&\frac{\mu^{\gamma}_{t}([\ln\frac{V_{T}}{V_{t}}]^{+})}{T}
=\liminf_{T\rightarrow\infty }\frac{\ln E[\exp(\gamma [\ln \frac{V_{T}}{V_{t}}]^{+})|\mathcal{F}_{t}]}{\gamma T}
=\liminf_{T\rightarrow\infty }\frac{1}{T}\frac{1}{\gamma}\ln E\Big[\max{(\frac{V_{T}}{V_{t}},1)}^{\gamma}|\mathcal{F}_{t}\Big]\notag\\
&=\liminf_{T\rightarrow\infty }\frac{1}{T}\frac{1}{\gamma}\ln E\Big[\frac{\max{(V_{T},V_{t})^{\gamma}}}{V_{t}^{\gamma}}|\mathcal{F}_{t}\Big]
=\liminf_{T\rightarrow\infty }\Big[\frac{1}{T}\frac{1}{\gamma}\ln E[\max{(V_{T},V_{t})}^{\gamma}|\mathcal{F}_{t}]-\frac{1}{T}\ln V_{t}\Big] \notag\\
&=\liminf_{T\rightarrow\infty }\frac{1}{T}\frac{1}{\gamma}\ln E[\max{(V_{T},V_{t})}^{\gamma}|\mathcal{F}_{t}]. \label{eq:9d}
\end{align}
Using the above, and the fact that $V_{T}\leq \max{(V_{T},V_{t})}$, and $\mu^{\gamma}_{t}([\ln\frac{V_{T}}{V_{t}}]^{+})\geq 0$, for all $V\in\mathbb{V}$, we have the following inequality
\begin{equation}\label{eq:9c}
\Big[\liminf_{T\rightarrow\infty }\frac{1}{T}\frac{1}{\gamma}\ln E[V_{T}^{\gamma}|\mathcal{F}_{t}]\Big]^{+}\leq \liminf_{T\rightarrow\infty }\frac{\mu^{\gamma}_{t}([\ln\frac{V_{T}}{V_{t}}]^{+})}{T}.
\end{equation}
Next, we will prove the converse inequality. Without loss of generality, using locality, and the fact that the function $[\cdot]^{+}$ is non-negative, we could assume that
\begin{equation}\label{eq:9a}
\liminf_{T\rightarrow\infty }\frac{\mu^{\gamma}_{t}([\ln\frac{V_{T}}{V_{t}}]^{+})}{T}>0.
\end{equation}
Let $X_{T}:=E[\1_{\{V_{T}>V_{t}\}}V_{T}^{\gamma}|\mathcal{F}_{t}]$. Using \eqref{eq:9c}, \eqref{eq:9d}, and because $E[\1_{\{V_{T}\leq V_{t}\}}V_{t}^{\gamma}|\mathcal{F}_{t}]\leq V_{t}^{\gamma}$, we get
\begin{align}
\liminf_{T\rightarrow\infty}\frac{1}{T}\frac{1}{\gamma}\ln X_{T} & \leq \Big[\liminf_{T\rightarrow\infty }\frac{1}{T}\frac{1}{\gamma}\ln E[V_{T}^{\gamma}|\mathcal{F}_{t}]\Big]^{+}\leq \liminf_{T\rightarrow\infty }\frac{\mu^{\gamma}_{t}([\ln\frac{V_{T}}{V_{t}}]^{+})}{T}\notag \\
& = \liminf_{T\rightarrow\infty }\frac{1}{T}\frac{1}{\gamma}\ln E[\max{(V_{T},V_{t})}^{\gamma}|\mathcal{F}_{t}] \leq\liminf_{T\rightarrow\infty }\frac{1}{T}\frac{1}{\gamma}\ln(X_{T}+V_{t}^{\gamma}).\label{eq:9b}
\end{align}
Due to \eqref{eq:9a}, and the fact that $\gamma>0$, we have $(X_{T}+V_{t}^{\gamma})\stackrel{T\rightarrow \infty}{\longrightarrow} \infty$, and consequently $X_{T}\stackrel{T\rightarrow \infty}{\longrightarrow} \infty$. Thus,
$$
|\ln(X_{T}+V_{t}^{\gamma}) -\ln(X_{T})|\rightarrow 0, \quad T\rightarrow\infty.
$$
Using \eqref{eq:9b} we conclude the proof.

\smallskip
\noindent 4) This is a direct result of the analogous property for negative of the dynamic monetary entropic utility. See Proposition~\ref{prop:entropicrisk}.

\smallskip
\noindent 5) Let $s\geq t\geq 0\in\mathbb{T}$, $V\in\widetilde{\mathbb{V}}$, and $m_{s}\in \bar{L}^{0}_{s}$. It is enough to prove that
\begin{equation}\label{eq:1p1}
e^{\varphi^{\gamma}_{s}(V)}\geq e^{m_{s}} \Rightarrow e^{\varphi^{\gamma}_{t}(V)}\geq e^{E[m_{s}|\mathcal{F}_{t}]}.
\end{equation}
It is easy to note, that
\begin{align*}
e^{\varphi^{\gamma}_{s}(V)} & =e^{\liminf_{T\rightarrow\infty}\frac{1}{T}\frac{1}{\gamma}\ln E[V_{T}^{\gamma}|\mathcal{F}_{s}]}=e^{\liminf_{T\rightarrow\infty}\ln \big[ E[V_{T}^{\gamma}|\mathcal{F}_{s}]^{\frac{1}{\gamma T}}\big]}\\
& =\liminf_{T\rightarrow\infty}e^{\ln \big[ E[V_{T}^{\gamma}|\mathcal{F}_{s}]^{\frac{1}{\gamma T}}\big]}=\liminf_{T\rightarrow\infty} E[V_{T}^{\gamma}|\mathcal{F}_{s}]^{\frac{1}{\gamma T}}.
\end{align*}
Using this, we conclude that~(\ref{eq:1p1}) is equivalent to the following
\begin{equation}\label{eq:1p2}
\liminf_{T\rightarrow\infty} E[V_{T}^{\gamma}|\mathcal{F}_{s}]^{\frac{1}{\gamma T}}\geq e^{m_{s}} \Rightarrow \liminf_{T\rightarrow\infty} E[V_{T}^{\gamma}|\mathcal{F}_{t}]^{\frac{1}{\gamma T}}\geq e^{E[m_{s}|\mathcal{F}_{t}]}.
\end{equation}
Assume that $\liminf_{T\rightarrow\infty} E[V_{T}^{\gamma}|\mathcal{F}_{s}]^{\frac{1}{\gamma T}}\geq e^{m_{s}}$. Due to the tower property we have
$$
\liminf_{T\rightarrow\infty} E[V_{T}^{\gamma}|\mathcal{F}_{t}]^{\frac{1}{\gamma T}}=\liminf_{T\rightarrow\infty} E\big[ E[V_{T}^{\gamma}|\mathcal{F}_{s}]|\mathcal{F}_{t}\big]^{\frac{1}{\gamma T}}.
$$
Since, $0<\frac{1}{\gamma T}<1$, for $T$ large enough, we get that the function $f(x)=x^{\frac{1}{\gamma T}},\  x>0$, is concave. Consequently, by Jensen's inequality, we continue
$$
\liminf_{T\rightarrow\infty}E\big[ E[V_{T}^{\gamma}|\mathcal{F}_{s}]|\mathcal{F}_{t}\big]^{\frac{1}{\gamma T}} \geq \liminf_{T\rightarrow\infty} E\big[ E[V_{T}^{\gamma}|\mathcal{F}_{s}]^{\frac{1}{\gamma T}}|\mathcal{F}_{t}\big].
$$
Since, $E[V_{T}^{\gamma}|\mathcal{F}_{s}]^{\frac{1}{\gamma T}}$ is non-negative for every $T\in\bT$, by Fatou lemma, we conclude
$$
\liminf_{T\rightarrow\infty} E\big[ E[V_{T}^{\gamma}|\mathcal{F}_{s}]^{\frac{1}{\gamma T}}|\mathcal{F}_{t}\big]\geq E\big[\liminf_{T\rightarrow\infty}  E[V_{T}^{\gamma}|\mathcal{F}_{s}]^{\frac{1}{\gamma T}}|\mathcal{F}_{t}\big].
$$
Finally, using the fact that $\liminf_{T\rightarrow\infty} E[V_{T}^{\gamma}|\mathcal{F}_{s}]^{\frac{1}{\gamma T}}\geq e^{m_{s}}$, and by Jensen's inequality for $f(x)=e^{x}$, we get
$$
E\big[\liminf_{T\rightarrow\infty}  E[V_{T}^{\gamma}|\mathcal{F}_{s}]^{\frac{1}{\gamma T}}|\mathcal{F}_{t}\big] \geq E[e^{m_{s}}|\mathcal{F}_{t}]\geq e^{E[m_{s}|\mathcal{F}_{t}]},
$$
which completes the proof.

\smallskip
\noindent 6) Let $t\in\mathbb{T}$, $V\in\widetilde{\mathbb{V}}$ and $\gamma<0$.  We want to prove that for $s\in\mathbb{T}$, $s>t,$ and $m_{s}\in \bar{L}^{0}_{s}$, we have
\begin{equation}\label{eq:2p1}
\varphi^{\gamma}_{s}(V)\leq m_{s} \Rightarrow \varphi^{\gamma}_{t}(V)\leq E[m_{s}|\mathcal{F}_{t}].
\end{equation}
Doing similar operations as in 5), we deduce that \eqref{eq:2p1} is equivalent to
\begin{equation}\label{eq:2p1-1}
\liminf_{T\rightarrow\infty} E[V_{T}^{\gamma}|\mathcal{F}_{s}]^{\frac{1}{\gamma T}}\leq e^{m_{s}} \Rightarrow \liminf_{T\rightarrow\infty} E[V_{T}^{\gamma}|\mathcal{F}_{t}]^{\frac{1}{\gamma T}}\leq e^{E[m_{s}|\mathcal{F}_{t}]}.
\end{equation}
Since for $\gamma<0$ and nonnegative $x$ the function $f(x)=x^{\gamma}$ is decreasing, we have that \eqref{eq:2p1-1} is equivalent to
$$
\Big[\liminf_{T\rightarrow\infty} E[V_{T}^{\gamma}|\mathcal{F}_{s}]^{\frac{1}{\gamma T}}\Big]^{\gamma} \geq e^{\gamma m_{s}}  \Rightarrow \Big[\liminf_{T\rightarrow\infty} E[V_{T}^{\gamma}|\mathcal{F}_{t}]^{\frac{1}{\gamma T}}\Big]^{\gamma} \geq e^{\gamma E[m_{s}|\mathcal{F}_{t}]}
$$
which is consequently equivalent to
$$
\limsup_{T\rightarrow\infty}\Big[E[V_{T}^{\gamma}|\mathcal{F}_{s}]^{\frac{1}{\gamma T}}\Big]^{\gamma} \geq e^{\gamma m_{s}}  \Rightarrow \limsup_{T\rightarrow\infty} \Big[E[V_{T}^{\gamma}|\mathcal{F}_{t}]^{\frac{1}{\gamma T}}\Big]^{\gamma} \geq e^{\gamma E[m_{s}|\mathcal{F}_{t}]},
$$
From here, we conclude that \eqref{eq:1p1} is equivalent to
\begin{equation}\label{eq:2p1-3}
\limsup_{T\rightarrow\infty}E[V_{T}^{\gamma}|\mathcal{F}_{s}]^{\frac{1}{T}} \geq e^{\gamma m_{s}}  \Rightarrow \limsup_{T\rightarrow\infty} E[V_{T}^{\gamma}|\mathcal{F}_{t}]^{\frac{1}{T}}\geq e^{\gamma E[m_{s}|\mathcal{F}_{t}]},
\end{equation}
and thus we will verify this implication.

To give a better intuition of the proof of \eqref{eq:2p1-3}, first we will consider $t=0$, i.e we will show that
that for any $m_{s}\in\bar{L}^{0}_{s}$,  we have that
\begin{equation}\label{eq:2p1-4}
\limsup_{T\rightarrow\infty}E[V_{T}^{\gamma}|\mathcal{F}_{s}]^{\frac{1}{T}} \geq e^{\gamma m_{s}} \Rightarrow \limsup_{T\rightarrow\infty} E[V_{T}^{\gamma}]^{\frac{1}{T}}\geq e^{\gamma E[m_{s}]}.
\end{equation}
Assume that $s>0$,  $m_{s}\in \bar{L}^{0}_{s}$, and such that
$$
\limsup_{T\rightarrow\infty}E[V_{T}^{\gamma}|\mathcal{F}_{s}]^{\frac{1}{T}} \geq e^{\gamma m_{s}}.
$$
Note that, there exists a set $C\in\mathcal{F}_{s}$, such that $P[C]>0$ and $\1_{C}e^{\gamma m_{s}} \geq \1_{C}E[e^{\gamma m_{s}}]$.
Hence,
$$
\1_{C}\limsup_{T\rightarrow\infty}E[V_{T}^{\gamma}|\mathcal{F}_{s}]^{\frac{1}{T}} \geq \1_{C}E[e^{\gamma m_{s}}].
$$
By Jensen's inequality, we continue
\begin{equation}\label{eq:2p2}
\1_{C}\limsup_{T\rightarrow\infty}E[V_{T}^{\gamma}|\mathcal{F}_{s}]^{\frac{1}{T}} \geq \1_{C}e^{\gamma E[m_{s}]}.
\end{equation}
Let $\epsilon>0$,  and  put $B^{\epsilon}_{T}:=\{\omega\in\Omega: E[V_{T}^{\gamma}|\mathcal{F}_{s}]^{\frac{1}{T}}(\omega) \geq e^{\gamma E[m_{s}]-\epsilon}\big\}$.
Notice that
$$
C \subset \limsup_{T\rightarrow\infty} B_{T}^{\epsilon},
$$
which consequently implies that
\begin{equation}\label{eq:2p3}
P\Big[\limsup_{T\rightarrow\infty} B_{T}^{\epsilon}\Big] > 0.
\end{equation}
From here, by Borel-Cantelli Lemma, we get that $\sum_{T=1}^{\infty}P[B_{T}^{\epsilon}]=\infty$.
Since the last series is divergent, there exists a subsequence $\{T^{\epsilon}_{k}\}_{(k=1,2,\ldots)}$ such that
$$
P[B_{T^{\epsilon}_{k}}^{\epsilon}]\geq\frac{1}{(T^{\epsilon}_{k})^{2}}.
$$
Using this, we have the following chain of inequalities
\begin{align*}
\limsup_{T\rightarrow\infty} E[V_{T}^{\gamma}]^{\frac{1}{T}} & =\limsup_{T\rightarrow\infty} E[E[V_{T}^{\gamma}|\mathcal{F}_{s}]]^{\frac{1}{T}}\geq \limsup_{T\rightarrow\infty} E[\1_{B_{T}^{\epsilon}}E[V_{T}^{\gamma}|\mathcal{F}_{s}]]^{\frac{1}{T}}\\
& \geq \limsup_{T\rightarrow\infty} E[\1_{B_{T}^{\epsilon}}e^{(\gamma E[m_{s}]-\epsilon)T}]^{\frac{1}{T}}\geq e^{\gamma E[m_{s}]-\epsilon}\limsup_{T\rightarrow\infty} P[B_{T}^{\epsilon}]^{\frac{1}{T}}\\
& \geq e^{\gamma E[m_{s}]-\epsilon}\limsup_{T_{k}^{\epsilon}\rightarrow\infty} P[B_{T_{k}^{\epsilon}}^{\epsilon}]^{\frac{1}{T_{k}^{\epsilon}}} \geq e^{\gamma E[m_{s}]-\epsilon}\limsup_{T^{\epsilon}_{k}\rightarrow\infty} \Big[\frac{1}{(T^{\epsilon}_{k})^{2}}\Big]^{\frac{1}{T^{\epsilon}_{k}}}\\
& = e^{\gamma E[m_{s}]-\epsilon}.
\end{align*}
Hence, taking into account that $\epsilon>0$ was arbitrary chosen, implication \eqref{eq:2p1-4} follows immediately.

The proof for $t>0$ follows similar line of ideas as for $t=0$, although it is a bit more technical.
For sake of completeness we will present the proof here too.
The proof is done by contradiction:  assume that~\eqref{eq:2p1} is not true for some $s\in\mathbb{T}$, $s>t$ .
Then, since \eqref{eq:2p1} is equivalent to \eqref{eq:2p1-3}, there exists $V\in\tilde{\mathbb{V}}$, $m_{s}\in \bar{L}_{s}^{0}$ and $A\in\mathcal{F}_{t}$, $P[A]>0$ such that for
\begin{equation}\label{eq:2p4}
\limsup_{T\rightarrow\infty}E[V_{T}^{\gamma}|\mathcal{F}_{s}]^{\frac{1}{T}} \geq e^{\gamma m_{s}} \quad\textrm{and}\quad \limsup_{T\rightarrow\infty} E[V_{T}^{\gamma}|\mathcal{F}_{t}]^{\frac{1}{T}}< e^{\gamma E[m_{s}|\mathcal{F}_{t}]}.
\end{equation}
almost surely on $A$.
Note that there exists $\epsilon>0$ and $A_{2}\in\mathcal{F}_{t}$, $A_{2}\subset A$, $P[A_{2}]>0$, such that
\begin{equation}\label{eq:2p5}
1_{A_{2}}\limsup_{T\rightarrow\infty} E[V_{T}^{\gamma}|\mathcal{F}_{t}]^{\frac{1}{T}}\leq 1_{A_{2}}e^{\gamma E[m_{s}|\mathcal{F}_{t}]-2\epsilon}.
\end{equation}
Let us consider the following sets
\begin{align*}
B^{\epsilon}_{T} & :=\{\omega\in A_{2}: E[V_{T}^{\gamma}|\mathcal{F}_{s}]^{\frac{1}{T}}(\omega) \geq e^{\gamma E[m_{s}|\mathcal{F}_{t}](\omega)-\epsilon}\big\},\\
D_{\alpha} & :=\{\omega\in A_{2}: \sum_{T=1}^{\infty} E[1_{B_{T}^{\epsilon}}|\mathcal{F}_{t}]<\alpha\},\quad\quad \alpha\in \bN \cup\set{+\infty}.
\end{align*}
Note that $D_n\in\cF_t$ for any $n\in\bN$, $D_n\subset D_m$ for $n\leq m$, and $D_\infty = \cup_{n\in\bN}D_n\in\cF_t$.
Next we consider two cases: a) $P[D_{\infty}]>0$ and b) $P[D_{\infty}]=0$.

\smallskip
\noindent  Case a) Since $P[D_{\infty}] = P[\lim_{n\to\infty} D_n] =\lim_{n\to\infty} P[D_n] >0$, there exists $n_{0}>0$ such that $P[D_{n_{0}}]>0$.
Consequently,
$$
\sum_{T=1}^{\infty}P[B_{T}^{\epsilon}\cap D_{n_{0}}]<n_{0}.
$$
From here, by Borel-Cantelli Lemma, we get
$$
P\Big[\limsup_{T\rightarrow\infty} [B_{T}^{\epsilon}\cap D_{n_{0}}]\Big]=0,
$$
which implies that
$$
1_{D_{n_{0}}}\limsup_{T\rightarrow\infty}E[V_{T}^{\gamma}|\mathcal{F}_{s}]^{\frac{1}{T}} \leq 1_{D_{n_{0}}}e^{\gamma E[m_{s}|\mathcal{F}_{t}]-\epsilon}.
$$
that contradicts~\eqref{eq:2p4} on some set of positive measure.

\smallskip
\noindent  Case b) Let $P[D_{\infty}]=0$. First note that,
\begin{align}
\limsup_{T\rightarrow\infty} E[V_{T}^{\gamma}|\mathcal{F}_{t}]^{\frac{1}{T}}=\limsup_{T\rightarrow\infty} E[E[V_{T}^{\gamma}|\mathcal{F}_{s}]|\mathcal{F}_{t}]^{\frac{1}{T}}\geq \limsup_{T\rightarrow\infty} E[\1_{B_{T}^{\epsilon}}E[V_{T}^{\gamma}|\mathcal{F}_{s}]|\mathcal{F}_{t}]^{\frac{1}{T}} \notag \\
\geq \limsup_{T\rightarrow\infty} E[\1_{B_{T}^{\epsilon}}e^{(\gamma E[ m_{s}|\mathcal{F}_{t}]-\epsilon)T}|\mathcal{F}_{t}]^{\frac{1}{T}}\geq e^{\gamma E[m_{s}|\mathcal{F}_{t}]-\epsilon}\limsup_{T\rightarrow\infty} E[1_{B_{T}^{\epsilon}}|\mathcal{F}_{t}]^{\frac{1}{T}}. \label{eq:2p1-5}
\end{align}
Since $D_\infty\subset A_2$, and $P[D_{\infty}]=0$, we have that for (almost) every $\omega\in A_{2}$ there exists a subsequence  $\{ T^{\epsilon,\omega}_{k}\}_{k\in\bN}$ such that
$$
E\big[1_{B_{ T^{\epsilon,\omega}_{k}}^{\epsilon}}|\mathcal{F}_{t}\big](\omega)\geq\frac{1}{(T^{\epsilon,\omega}_{k})^{2}}.
$$
Using this, and \eqref{eq:2p1-5}, we conclude that for (almost) every $\omega\in A_{2}$
$$
\limsup_{T\rightarrow\infty} E[1_{B_{T}^{\epsilon}}|\mathcal{F}_{t}]^{\frac{1}{T}}(\omega)
\geq
\limsup_{ T^{\epsilon,\omega}_{k}\rightarrow\infty} E[1_{B_{ T^{\epsilon,\omega}_{k}}^{\epsilon}}|\mathcal{F}_{t}]^{\frac{1}{ T^{\epsilon,\omega}_{k}}}(\omega)
\geq
\limsup_{ T^{\epsilon,\omega}_{k}\rightarrow\infty}\Big[\frac{1}{
(T^{\epsilon,\omega}_{k})^{2}}\Big]^{\frac{1}{ T^{\epsilon,\omega}_{k}}}=1.
$$
Thus, almost everywhere on $A_{2}$
$$
\limsup_{T\rightarrow\infty} E[V_{T}^{\gamma}|\mathcal{F}_{t}]^{\frac{1}{T}}\geq e^{\gamma E[m_{s}|\mathcal{F}_{t}]-\epsilon}.
$$
Combining the last inequality with~\eqref{eq:2p5}, we get
$$
1_{A_{2}}e^{\gamma E[m_{s}|\mathcal{F}_{t}]-2\epsilon}\geq 1_{A_{2}}\limsup_{T\rightarrow\infty} E[V_{T}^{\gamma}|\mathcal{F}_{t}]^{\frac{1}{T}}\geq 1_{A_{2}}e^{\gamma E[m_{s}|\mathcal{F}_{t}]-\epsilon},
$$
which leads to contradiction, as $P[A_{2}]>0$.
\end{proof}

\begin{proof}[\textbf{Proof of Proposition~\ref{pr:DRSCadd}}]
Let $([0,1],\mathcal{B}([0,1]),\{\mathcal{F}_{t}\}_{t\in\mathbb{N}_{0}},\lambda)$ be a filtered probability space, with $\cF_{0}=\{[0,1],\emptyset\}$ and $\cF_{1}=\mathcal{B}([0,1])$.

\smallskip
\noindent 3') For $\gamma=-1$ it is enough to consider a simple example
$$
 \widehat V_{T}(\omega)=\left\{
\begin{array}{ll}
e^{-T} & \omega\in [0,e^{-T}],\\
e^{T} & \omega\in [e^{-T},1].
\end{array}\right.
$$
This example could be easily modified for any $\gamma<0$. For $\gamma=0$ it is enough to consider
$$
 \widehat V'_{T}(\omega)=\left\{
\begin{array}{ll}
e^{-T^{2}} & \omega\in [0,\frac{1}{T}],\\
e^{T} & \omega\in [\frac{1}{T},1].
\end{array}\right.
$$

\smallskip
\noindent 5') Let $\gamma=1$, and let $\{ \widehat V_{T}\}_{T\in \mathbb{N}}$ be defined by\begin{equation}\label{eq:notAcc}
 \widehat V_{T}(\omega)=\left\{
\begin{array}{ll}
\frac{1}{T} & \omega\in [0,\frac{1}{T}],\\
e^{T} & \omega\in [\frac{1}{T},1].
\end{array}\right.
\end{equation}
For $\omega\neq 0$, we have
$$
\varphi^{-1}_{1}( \widehat V_{T})(\omega)= \liminf_{T\rightarrow\infty}\frac{-1}{T}\ln \frac{1}{ \widehat V_{T}(\omega)}=\liminf_{T\rightarrow\infty}[(-\frac{\ln T}{T})\cdot\1_{[0,\frac{1}{T}]}(\omega)+1\cdot\1_{[\frac{1}{T},1]}(\omega)]=1.
$$
On the other hand
$$
\varphi^{-1}_{0}( \widehat V_{T})=\liminf_{T\rightarrow\infty}\frac{-1}{T}\ln E(\frac{1}{ \widehat V_{T}})=\liminf_{T\rightarrow\infty}\frac{-1}{T}\ln (1+\frac{T-1}{T}e^{-T})\leq \liminf_{T\rightarrow\infty}\frac{-\ln 1}{T}=0.
$$
Thus, with $m_{1}=1$, we get
$$
\varphi^{-1}_{1}( \widehat V)\geq m_1 \not\Rightarrow \varphi^{-1}_{0}( \widehat V)\geq E[m_1|\cF_{0}],
$$
which contradicts supermartingale consistency. This counterexample can be easily adjusted for any $\gamma<0$.

Similarly, for $\gamma=0$, we consider
$$
\widehat V'_{T}(\omega):=\left\{
\begin{array}{ll}
e^{-T^{2}} & \omega\in [0,\frac{1}{T}],\\
e^{T} & \omega\in [\frac{1}{T},1].
\end{array}\right.
$$

\smallskip
\noindent 6') As in the previous case we will consider only $\gamma=1$ and $\gamma=0$.
For $\gamma=1$, we take $\{ \widehat V_{T}\}_{T\in \mathbb{N}}$ defined by
\begin{equation}\label{eq:notRej}
 \widehat V_{T}(\omega)=\left\{
\begin{array}{ll}
Te^{T} & \omega\in [0,\frac{1}{T}],\\
1 & \omega\in [\frac{1}{T},1].
\end{array}\right.
\end{equation}
Then, we have
$$
\varphi^{1}_{1}( \widehat V_{T})(\omega)= \liminf_{T\rightarrow\infty}\frac{1}{T}\ln  \widehat V_{T(\omega)}=\liminf_{T\rightarrow\infty}[(1+\frac{\ln T}{T})\cdot\1_{[0,\frac{1}{T}]}(\omega)+0\cdot\1_{[\frac{1}{T},1]}(\omega)]=0,\quad \omega\neq 0.
$$
On the other hand
$$
\varphi^{1}_{0}( \widehat V_{T})=\liminf_{T\rightarrow\infty}\frac{1}{T}\ln E( \widehat V_{T})=\liminf_{T\rightarrow\infty}\frac{1}{T}\ln (e^{T}+\frac{T-1}{T})\geq \liminf_{T\rightarrow\infty}\frac{T}{T}=1.
$$
Thus, with $m_{1}=0$, we get
$$
\varphi^{1}_{1}( \widehat V)\leq m_1 \not\Rightarrow \varphi^{1}_{0}( \widehat V)\leq E[m_1|\cF_{0}],
$$
which contradicts submartingale consistency.

Similarly, for $\gamma=0$, we consider
$$
 \widehat V'_{T}(\omega)=\left\{
\begin{array}{ll}
e^{T^{2}} & \omega\in [0,\frac{1}{T}],\\
1 & \omega\in [\frac{1}{T},1].
\end{array}\right.
$$
\end{proof}

\end{appendix}

\section*{Acknowledgments}
Tomasz R. Bielecki and Igor Cialenco acknowledge support from the NSF grant DMS-0908099, and DMS-1211256.
Marcin Pitera acknowledges the support by Project operated within the Foundation for Polish Science IPP Programme "Geometry and Topology in Physical Models" co-financed by the EU European Regional Development Fund, Operational Program Innovative Economy 2007-2013.

\bibliographystyle{amsplain}
%\bibliography{bibliografia}

\begin{thebibliography}{10}

\bibitem{AccPen2010}
B.~Acciaio and I.~Penner, \emph{Dynamic risk measures}, In G.Di Nunno and B.
  {\"O}ksendal (Eds.) Advanced Mathematical Methods for Finance (2011), 1--34.

\bibitem{AraBorFerGhoMar1993}
A.~Arapostathis, V.~S. Borkar, E.~Fern{\'a}ndez-Gaucherand, M.~K. Ghosh, and
  S.~I. Marcus, \emph{Discrete-time controlled markov processes with average
  cost criterion: a survey}, SIAM Journal on Control and Optimization
  \textbf{31} (1993), no.~2, 282--344.

\bibitem{BiaFri2010}
S.~Biagini and M.~Frittelli, \emph{On the extension of the {N}amioka-{K}lee
  theorem and on the {F}atou property for risk measures}, Optimality and Risk -
  Modern Trends in Mathematical Finance, Springer Berlin Heidelberg, 2010,
  pp.~1--28 (English).

\bibitem{BieCiaDraKar2013}
T.~R. Bielecki, I.~Cialenco, S.~Drapeau, and M.~Karliczek, \emph{{Dynamic
  Assessment Indices}}, ArXiv e-prints (2013).

\bibitem{BieCiaZha2012}
T.~R. Bielecki, I.~Cialenco, and Z.~Zhang, \emph{Dynamic coherent acceptability
  indices and their applications to finance}, Mathematical Finance (2013).

\bibitem{BiePli1999}
T.~R. Bielecki and S.~R. Pliska, \emph{Risk-sensitive dynamic asset
  management}, Appl. Math. Optim. \textbf{39} (1999), no.~3, 337--360.

\bibitem{BiePli2003}
\bysame, \emph{Economic properties of the risk sensitive criterion for
  portfolio management}, Review of Accounting and Finance \textbf{2} (2003),
  3--17.

\bibitem{CheDelKup2006}
P.~Cheridito, F.~Delbaen and M.~Kupper, \emph{Dynamic monetary risk measures for bounded discrete-time processes}, Electronic Journal of Probability \textbf{11} (2006), no.~3, 57--106.

\bibitem{CheKup2009}
P.~Cheridito and M.~Kupper, \emph{Recursiveness of indifference prices and
  translation-invariant preferences}, Mathematics and Financial Economics
  \textbf{2} (2009), no.~3, 173--188.

\bibitem{CheLi2009}
P.~Cheridito and T.~Li, \emph{Risk measures on {O}rlicz hearts}, Math. Finance
  \textbf{19} (2009), no.~2, 189--214.

\bibitem{CheMad2009}
A.~S. Cherny and D.~B. Madan, \emph{New measures for performance evaluation},
  The Review of Financial Studies \textbf{22} (2009), no.~7, 2571--2606.

\bibitem{DetSca2005}
K.~Detlefsen and G.~Scandolo, \emph{Conditional and dynamic convex risk
  measures}, Finance and Stochastics \textbf{9} (2005), no.~4, 539--561
  (English).

\bibitem{DiMSte2007}
G.~Di~Masi and \L{}. Stettner, \emph{Infinite horizon risk sensitive control of
  discrete time markov processes under minorization property}, SIAM Journal on
  Control and Optimization \textbf{46} (2007), no.~1, 231--252.

\bibitem{FilKupVog2009}
D.~Filipovi{\'c}, M.~Kupper, and N.~Vogelpoth, \emph{Separation and duality in
  locally -convex modules}, Journal of Functional Analysis \textbf{256} (2009),
  no.~12, 3996 -- 4029.

\bibitem{FleShe2000}
W.~H. Fleming and S.~J. Sheu, \emph{Risk-sensitive control and an optimal
  investment model}, Mathematical Finance \textbf{10} (2000), no.~2, 197--213.

\bibitem{KupSch2009}
M.~Kupper and W.~Schachermayer, \emph{Representation results for law invariant
  time consistent functions}, Mathematics and Financial Economics \textbf{2}
  (2009), no.~3, 189--210.

\bibitem{Whi1990}
P.~Whittle, \emph{Risk-sensitive optimal control}, Wiley New York, 1990.


\end{thebibliography}

\providecommand{\bysame}{\leavevmode\hbox to3em{\hrulefill}\thinspace}
\providecommand{\MR}{\relax\ifhmode\unskip\space\fi MR }
\providecommand{\MRhref}[2]{%
  \href{http://www.ams.org/mathscinet-getitem?mr=#1}{#2}
}
\providecommand{\href}[2]{#2}

 \end{document}